\documentclass[10pt]{article}

\newcommand{\blind}{1}

\usepackage{scalerel}
\usepackage{listings}
\usepackage{cancel}
\usepackage[normalem]{ulem}
\usepackage{comment}
\usepackage{float}
\usepackage{array,multirow}
\usepackage{placeins}
\usepackage{epsfig}
\usepackage[english]{babel}
\usepackage{graphicx}
\usepackage{setspace}
\usepackage{amsmath}
\usepackage{relsize}
\usepackage{color,soul}
\usepackage{makeidx}
\usepackage[caption = false]{subfig}
\usepackage{multicol}
\usepackage{apacite}
\usepackage{setspace}

\usepackage[super,sort&compress]{natbib}
\setcitestyle{comma,numbers,super,open={},close={}} 
\makeatletter
\renewcommand\@biblabel[1]{#1.}
\makeatother

\begin{document}
	
	\def\spacingset#1{\renewcommand{\baselinestretch}%
		{#1}\small\normalsize} \spacingset{1.1}
	
	\if1\blind
	{
		\title{\bf  Tolerance and Prediction Intervals\\
for Non-normal Models
}
		\author{Geoffrey S Johnson\vspace{1mm}\\		
			Merck \& Co., Inc. \\
			770 Sumneytown Pike, West Point, PA 19438 USA\\
			geoffrey.s.johnson@gmail.com\\
			     }
		\date{}
		\maketitle
	} \fi
	
	\if0\blind
	{
		\title{\bf Tolerance and Prediction Intervals}
		\date{}		
		\maketitle
	} \fi

	\title{}

	\label{firstpage}

	\begin{abstract}
A prediction interval covers a future observation from a random process in repeated sampling, and is typically constructed by identifying a pivotal quantity that is also an ancillary statistic.  Analogously, a tolerance interval covers a population percentile in repeated sampling and is often based on a pivotal quantity.  One approach we consider in non-normal models leverages a link function resulting in a pivotal quantity that is approximately normally distributed.  In settings where this normal approximation does not hold we consider a second approach for tolerance and prediction based on a confidence interval for the mean.  These methods are intuitive, simple to implement, have proper operating characteristics, and are computationally efficient compared to Bayesian, re-sampling, and machine learning methods.  This is demonstrated in the context of multi-site clinical trial recruitment with staggered site initiation, real-world time on treatment, and end-of-study success for a clinical endpoint.

	\end{abstract}
	
	\vfill
	
	\pagebreak
\section{Introduction}

A prediction interval covers a future observation from a random process, and is typically constructed by identifying a pivotal quantity that is also an ancillary statistic. Outside of normality it can sometimes be challenging to identify an ancillary pivotal quantity without assuming some of the model parameters are known. A common solution is to identify an appropriate transformation of the data that yields normally distributed observations.\cite{krishnamoorthy2008} Another is to treat model parameters as random variables and construct a Bayesian predictive distribution,\cite{anisimov2007, bani2013} analogous to bootstrapping a prediction interval.  Likewise, a tolerance interval covers a population percentile in repeated sampling and poses similar challenges outside of normality. The approach we consider leverages a link function resulting in a pivotal quantity that is approximately normally distributed and produces tolerance and prediction intervals that work well for non-normal models where identifying an exact pivotal quantity may be intractable. In settings where this normal approximation does not hold we consider a second approach for tolerance and prediction based on a confidence interval for the mean.  These methods are explored when modeling recruitment interarrival time in clinical trials, and ultimately, time to complete recruitment. 
\\

A key element when conducting a multicenter or multi-site clinical trial is modeling and predicting recruitment. This problem has been framed as a Poisson process where the observable site-level recruitment rate follows a Poisson distribution, resulting in site-level interarrival time that follows an exponential distribution.\cite{cox1980, cox1955, anisimov2007two, rukhin2006, senn1998} Under this framework Anisimov and Fedorov\cite{anisimov2007} treat the Poisson rate parameter itself as a gamma distributed random variable and plug parameter estimates into this doubly stochastic process or update a Bayesian prior predictive distribution to ultimately make study-level predictions for time to complete recruitment. Here we instead consider the modeling of interarrival time at the study level to avoid making site-level distributional assumptions and to allow greater flexibility in model selection. We also explore prediction intervals for time to complete recruitment that avoid making distributional assumptions about model parameters. This reduces overall model complexity while producing prediction intervals that maintain their nominal coverage probability. If site-level recruitment predictions are required the prediction intervals explored here can be adjusted or stratified to individual sites.  Moreover, the approach we consider is applicable even with staggered site initiation and a time-varying mean recruitment rate. 
\\

Another area of application for tolerance and prediction is 
time on treatment or time to treatment discontinuation in a real-world setting for a target patient population in a particular disease area.  This information is used in combination with economic valuations to support submissions to payers and for commercial forecasting.   
We apply the methods explored for tolerance and prediction to fully characterize time on treatment, and demonstrate how to summarize these analyses in a single visualization so that stakeholders can quickly and easily identify the required results for decision making.  The topic of prediction has also been promoted in the literature for clinical endpoints in phases 2 and 3 of pharmaceutical development.  Under a Bayesian framework many authors and practitioners interpret predictive inference as a legitimate probability statement regarding end-of-study success, and argue that this should supersede a typical power calculation.  Adopting an objective definition of probability we instead consider predictive inference as a statement of confidence based on a p-value unconditional on model parameters.  Just as with clinical trial recruitment, the intervals we explore to predict end-of-study success for a clinical endpoint do not require distributional assumptions on model parameters, reducing model and computational complexity while maintaining performance.  
\\

The novelty of this manuscript is on the interpretation and visualization of tolerance and prediction intervals using confidence curves, the statistical and computational evaluation of the methods explored in comparison to other methods for prediction, and the application of these methods to address recruitment, time on treatment, and end-of-study success.  Section \ref{development} presents the rationale for the intervals we consider and our interpretation will involve a blend of evidential p-values and a long-run error rate. Section \ref{simulation} explores their performance through simulation under a gamma data process and a doubly stochastic Poisson-gamma data process. Section \ref{application} provides a motivating example in the context of clinical trial recruitment and offers prediction limits easily implemented by a practitioner with limited knowledge of statistics. This section also provides methods to address staggered site initiation and a time-varying recruitment rate.  
Sections \ref{tot} and \ref{application success} present additional examples predicting time on treatment as well as end-of-study success for a clinical endpoint.  
Section \ref{closing remarks} provides closing remarks. SAS code is given in Appendix \ref{sas code}.

\section{Methodology}\label{development}
\subsection{Normality}
In repeated sampling a prediction interval covers a future observation of a random process $100(1-\alpha)\%$ of the time. For normally distributed data $\boldsymbol{Y}_n=Y_1,...,Y_n$ when the population variance $\sigma^2$ is known the pivotal quantity $(\bar{Y}_n-Y_{n+1})/\sigma\sqrt{1/n+1}$ is ancillary since it and its sampling distribution, $N(0,1)$, do not depend on the unknown mean $\mu$, where $\bar{Y}_n=(1/n)\sum_{i=1}^n Y_i$. When pivoted this quantity results in the interval estimate $\bar{y}_n\pm z_{1-\alpha/2}\cdot\sigma\sqrt{1/n+1}$, where $z_{1-\alpha/2}$ is the $100(1-\alpha/2)^{th}$ percentile of the standard normal distribution. This is a prediction interval for the as of yet unobserved $y_{n+1}$. When $\sigma^2$ is not known the ancillary pivotal quantity of choice becomes $(\bar{Y}_n-Y_{n+1})/S\sqrt{1/n+1}\sim T_{n-1}$, where $S^2$ is the bias corrected sample variance.  
In repeated sampling $\bar{y}_n\pm t_{n-1,1-\alpha/2}\cdot s\sqrt{1/n+1}$ will cover the $n+1^{th}$ observation $100(1-\alpha)\%$ of the time, regardless of the unknown fixed true $\mu$ and $\sigma^2$. The p-value testing the hypothesis $H_0$: $y_{n+1} \le c$ is given by $P\big(T_{n-1}\ge (\bar{y}_n-c)/s\sqrt{1/n+1}\big)$, the probability of the difference (discrepancy) between the observed result and the hypothesized future result or something more extreme, if these share unknown fixed parameters $\mu$ and $\sigma$. 
This probability forms the level of confidence that $y_{n+1}$ will be less than or equal to $c$, and is useful for controlling the type I error rate $\alpha$ when predicting $y_{n+1}$. The upper p-value function\cite{johnson2021} of all upper-tailed predictive p-values as a function of the hypothesis being tested is $H(y_{n+1})=1-\Phi_{n-1}\big(\hspace{1mm}\big[\bar{y}_n- y_{n+1}\big]\big/s\sqrt{1/n+1}\hspace{1mm} \big)$, where $\Phi_{n-1}$ denotes the cdf of a $T_{n-1}$ random variable. The null value $c$ is replaced with $y_{n+1}$ to denote that this is a function of all possible hypotheses around $y_{n+1}$. One can analogously define $H^-(y_{n+1})$ as the function of all lower-tailed p-values. The corresponding prediction confidence curve defined as 
\begin{eqnarray}
C(y_{n+1})&\equiv& \left\{ \begin{array}{cc}
H(y_{n+1}) & \text{if } y_{n+1}\le\bar{y}_n \\
 &  \nonumber\\
 H^{-}(y_{n+1})  & \text{if }y_{n+1}\ge\bar{y}_n, \end{array}  \right.\nonumber
\end{eqnarray}
and prediction confidence density $h(y_{n+1})\equiv dH(y_{n+1})/dy_{n+1}$, depict p-values and prediction intervals of all levels for hypotheses around $y_{n+1}$.
\\

A tolerance interval covers a population percentile in repeated sampling and is often used in engineering and manufacturing for quality control. In the case of a normally distributed population with $100p^{th}$ percentile $q_p=\mu+z_p\sigma$, the quantity $(\bar{Y}_n-q_p)/S\sqrt{1/n}$ follows a non-central $T_{n-1}(\nu)$ distribution with non-centrality parameter $\nu=-z_p\sqrt{n}$. This pivotal quantity yields $\big(\bar{y}_n+ t_{n-1,\alpha/2}(-\nu)\cdot s/\sqrt{n}\hspace{2mm},\hspace{2mm}\bar{y}_n+ t_{n-1,1-\alpha/2}(-\nu)\cdot s/\sqrt{n}\big)$ as a two-sided $100(1-\alpha)\%$ tolerance interval for $q_p$ of $N(\mu,\sigma^2)$. In repeated sampling this tolerance interval will cover the $100p^{th}$ percentile of the population $100(1-\alpha)\%$ of the time. The p-value testing the hypothesis $H_0$: $q_p \le c$ is given by $P\big(T_{n-1}(\nu)\ge (\bar{y}_n-c)/s\sqrt{1/n}\big)$, the probability of the observed result or something more extreme if the hypothesis for $q_p$ is true. This probability forms the level of confidence that $q_p$ is less than or equal to $c$, and is useful for controlling the type I error rate $\alpha$ when drawing conclusions about $q_p$. The two-sided $100(1-\alpha)\%$ tolerance interval that covers the middle $100p\%$ of the population is $\big(\bar{y}_n+t_{n-1,\alpha/2}(z_{(1-p)/2}\sqrt{n})\cdot s/\sqrt{n}\hspace{2mm}, \hspace{2mm}\bar{y}_n+ t_{n-1,1-\alpha/2}(z_{(1+p)/2}\sqrt{n})\cdot s/\sqrt{n}\big)$.
\\

Next we will explore candidates for approximate tolerance and prediction limits in the normal setting.  This will help guide our thinking in non-normal settings where exact pivotal quantities are not available or are difficult to obtain.  Notice the $100p^{th}$ percentile presented earlier contains the quantities $\mu$ and $\sigma$.  
We therefore consider using confidence limits for $\mu$ and $\sigma$ to construct an approximate upper tolerance limit as $\hat{q}_{p}^u=\hat{\mu}_{n}^u+z_{p}\hat{\sigma}_{n}^u=\bar{y}_n+t_{n-1,1-\alpha}\cdot s/\sqrt{n}+z_{p}\cdot \hat{\sigma}_{n}^u$, where $\hat{\mu}_n^u$ and $\hat{\sigma}_n^u$ are one-sided upper $100(1-\alpha)\%$ confidence limits for $\mu$ and $\sigma$ respectively.  This approximate tolerance limit is the $100p^{th}$ percentile of $N(\hat{\mu}_n^u,\hat{\sigma}_n^u)$.  To obtain an approximate two-sided tolerance interval that will cover the middle $100p\%$ of the population we consider using two-sided $100(1-\alpha)\%$ confidence limits for $\mu$ and $\sigma$ to construct $(\hat{\mu}_n^l+z_{(1-p)/2}\hat{\sigma}_n^u, \hspace{2mm}\hat{\mu}_n^u+z_{(1+p)/2}\hat{\sigma}_n^u)$.  These tolerance limits are the $100(1-p)/2^{th}$ and $100(1+p)/2^{th}$ percentiles of $N(\hat{\mu}_n^l,\hat{\sigma}_n^u)$ and $N(\hat{\mu}_n^u,\hat{\sigma}_n^u)$ respectively.    
\\

Prediction limits are closely related to tolerance limits.  Indeed, if we set $100p\%=100(1-\alpha)\%$ a two-sided tolerance interval for the middle $100p\%$ of the population is in a sense a conservative two-sided prediction interval.  If, however, we use $s$ in place of $\hat{\sigma}_n^u$ our approximate two-sided tolerance limits become $(\bar{y}_n-t_{n-1,1-\alpha/2}\cdot s/\sqrt{n}+z_{\alpha/2}\cdot s$, \hspace{2mm}$\bar{y}_n+t_{n-1,1-\alpha/2}\cdot s/\sqrt{n}+z_{1-\alpha/2}\cdot s)$ = $(\hat{\mu}_n^l+z_{\alpha/2}\cdot s,\hspace{2mm}\hat{\mu}_n^u+z_{1-\alpha/2}\cdot s)$.  These limits are the $100(\alpha/2)^{th}$ and $100(1-\alpha/2)^{th}$ percentiles of $N(\hat{\mu}_n^l,s)$ and $N(\hat{\mu}_n^u,s)$ respectively, and are asymptotically equivalent to $\bar{y}_n\pm t_{n-1,1-\alpha/2}\cdot s(1/\sqrt{n}+1)$ as $t_{n-1,1-\alpha/2}$ approaches $z_{1-\alpha/2}$ with increasing $n$.  Since $\underset{n\rightarrow\infty}{lim}(1/\sqrt{n}+1)$ = $\underset{n\rightarrow\infty}{lim}\sqrt{1/n+1}$ = 1 these limits are asymptotically equivalent to the prediction limits presented earlier, and since $(1/\sqrt{n}+1)$ $>$ $\sqrt{1/n+1}$ for $n\ge1$ they are conservative.  

\subsection{Non-normality}
Following the above discussion, we consider the general setting where $Y_1,...,Y_n,...,Y_N$ follow a cumulative distribution function F$(y;\mu,k)$ with location parameter $\mu$ and scale parameter $k$. 
Based on $n$ observations, our aim is to predict $\sum_{i=n+1}^{N} Y_i=\sum_{i=n+1}^{N} y_i$ with cumulative distribution function F$_{\scaleto{\sum Y}{5pt}}(y;\mu,k)$ for the data elements $\boldsymbol{Y}_{N-n}=Y_{n+1},...,Y_N$. 
When an exact ancillary pivotal quantity is not available or difficult to obtain we explore 
an intuitive pivotal quantity 
using estimators $\hat{\mu}(\boldsymbol{Y}_n)$ and $\hat{\mu}(\boldsymbol{Y}_{N-n})$ with link function $g\{\cdot\}$.  For ease of notation we will use $\hat{\mu}_n$ and $\hat{\mu}_{N-n}$ when referring to either an estimator or an estimate.  
Based on the asymptotic normality of  $g\{\hat{\mu}_n\}$ and $g\{\hat{\mu}_{N-n}\}$, $\big[g\{\hat{\mu}_n\}-g\{\hat{\mu}_{N-n}\}\big]\big/\hat{\text{SE}}_n
\overset{asymp}{\sim}T_{n-1}$, where $\hat{\text{SE}}_n$ is a model-based or empirical estimator for the standard error of $\big[g\{\hat{\mu}_n\}-g\{\hat{\mu}_{N-n}\}\big]$.  
 This quantity can be used to calculate approximate p-values and form prediction limits for hypotheses around $\sum_{i=n+1}^N y_i$, i.e. 
\begin{eqnarray}\label{option three}
(N-n)\cdot g^{-1}\Big\{g\{\hat{\mu}_n\}\pm t_{n-1,1-\alpha/2}\hat{\text{se}}_n\Big\}.
\label{option one}
\end{eqnarray}
When $g\{\cdot\}=\text{log}\{\cdot\}$, the asymptotic p-value testing $H_0$: $\sum_{i=n+1}^N y_i \le c$ is the probability of the ratio (discrepancy) between the observed result and the hypothesized future result or something more extreme. The upper p-value function is $H\big(\sum y\big)=1-\Phi_{n-1}\big(\hspace{1mm}\big[g\{\hat{\mu}_n\}-g\{\sum y/(N-n)\}\big]\big/\hat{\text{se}}_n
\hspace{1mm} \big)$, and the corresponding prediction confidence curve is defined as 
\begin{eqnarray}
C\big(\textstyle\sum y\big)&\equiv& \left\{ \begin{array}{cc}
H\big(\sum y\big) & \text{if } \sum y \le (N-n)\hat{\mu}_n\\
 &  \nonumber\\
 H^-\big(\sum y\big)  & \text{if }\sum y \ge (N-n)\hat{\mu}_n. \end{array}  \right.\nonumber
\end{eqnarray}
\\

When $g\{\hat{\mu}_{N-n}\}$ is not well approximated by a normal distribution we consider the intuitive prediction interval
\begin{eqnarray}
\Big(q_{\alpha/2}(\hat{\mu}_n^l,\hat{k}_n)\hspace{2mm}, \hspace{2mm}q_{1-\alpha/2}(\hat{\mu}_n^u,\hat{k}_n)\Big)\label{option two}
\end{eqnarray}
where $\hat{\mu}_n^l$ and $\hat{\mu}_n^u$ are lower and upper limits of a two-sided $100(1-\alpha)\%$ confidence interval for $\mu$, $\hat{k}_n$ is a point estimate for $k$, $q_{\alpha/2}(\hat{\mu}_n^l,\hat{k}_n)=\text{F}^{-1}_{\scaleto{\sum Y}{5pt}}(\alpha/2;\hat{\mu}_n^l,\hat{k}_n)$ is the $100(\alpha/2)^{th}$ percentile from F$_{\scaleto{\sum Y}{5pt}}(y;\hat{\mu}_n^l,\hat{k}_n)$
,  and $q_{1-\alpha/2}(\hat{\mu}_n^u,\hat{k}_n)=\text{F}^{-1}_{\scaleto{\sum Y}{5pt}}(1-\alpha/2;\hat{\mu}_n^u,\hat{k}_n)$ is the $100(1-\alpha/2)^{th}$ percentile from F$_{\scaleto{\sum Y}{5pt}}(y;\hat{\mu}_n^u,\hat{k}_n)$.  The p-value testing the hypothesis $H_0$: $\sum_{i=n+1}^{N} y_i\le c$ is the probability of the observed result, and that of the hypothesized future result, or something more extreme in opposite directions if these are equally probable results with shared $\mu$ and $k$.  
This is approximately equal to the probability of the difference or ratio (discrepancy) between the observed result and the hypothesized future result or something more extreme typically provided by an ancillary pivotal quantity. 
\\

Likewise, $q_p\big(\hat{\mu}_n,\hat{k}_n\big)$ is the parametric estimator for the $100p^{th}$ percentile of F$_{\scaleto{\sum Y}{5pt}}(y;\mu,k)$ based on estimators for $\mu$ and $k$. Based on the asymptotic normality of $g\{q_p\big(\hat{\mu}_n,\hat{k}_n\big)\}$ for some link function $g\{\cdot\}$, $\big[g\{q_p\big(\hat{\mu}_n,\hat{k}_n\big)\}-g\{q_p\}\big]\big/\hat{\text{SE}}_{n,p}\overset{asymp}{\sim}T_{n-1}$, where $\hat{\text{SE}}_{n,p}$ is a model-based or sandwich estimator for the standard error of $g\{q_p\big(\hat{\mu}_n,\hat{k}_n\big)\}$. An approximate two-sided $100(1-\alpha)\%$ tolerance interval that covers the middle $100p\%$ of F$_{\scaleto{\sum Y}{5pt}}(y;\mu,k)$ is given by 
\begin{eqnarray}
&&\Big(g^{-1}\Big\{g\{q_{(1-p)/2}(\hat{\mu}_n,\hat{k}_n)\}- t_{n-1,1-\alpha/2}\cdot\hat{\text{se}}_{n,(1-p)/2}\Big\}\hspace{2mm},\nonumber\\
&&\hspace{2mm} g^{-1}\Big\{g\{q_{(1+p)/2}(\hat{\mu}_n,\hat{k}_n)\}+t_{n-1,1-\alpha/2}\cdot\hat{\text{se}}_{n,(1+p)/2}\Big\}\Big).\label{option three}
\end{eqnarray}
Alternatively, when F$_{\scaleto{\sum Y}{5pt}}(y;\mu,k)$ is well approximated by a normal distribution, $\big[(N-n)\hat{\mu}_n-q_p\big]/(N-n)\hat{\text{SE}}_n$ follows a non-central $T_{n-1}(\nu)$ distribution with non-centrality parameter $\nu=-z_p\sqrt{n}/\sqrt{N-n}$, where $\hat{\text{SE}}_n$ is a model-based or sandwich estimator for the standard error of $\hat{\mu}_n$. An approximate two-sided tolerance interval that covers the middle $100p\%$ of F$_{\scaleto{\sum Y}{5pt}}(y;\mu,k)$ can also be constructed as 
\begin{eqnarray}
\big((N-n)\hat{\mu}_n+t_{n-1,\alpha/2}(z_{(1-p)/2}\sqrt{n}/\sqrt{N-n})\cdot (N-n)\hat{\text{se}}_n\hspace{2mm}\hspace{2.5mm},\nonumber\\ \hspace{2mm}(N-n)\hat{\mu}_n+ t_{n-1,1-\alpha/2}(z_{(1+p)/2}\sqrt{n}/\sqrt{N-n})\cdot (N-n)\hat{\text{se}}_n\big).\label{option four}
\end{eqnarray}
When neither F$_{\scaleto{\sum Y}{5pt}}(y;\mu,k)$ nor $g\{q_p\big(\hat{\mu}_n,\hat{k}_n\big)\}$ are well approximated by a normal distribution, we consider the approximate two-sided $100(1-\alpha)\%$ tolerance interval 
\begin{eqnarray}
\Big(q_{(1-p)/2}(\hat{\mu}_n^l,\hat{k}_n^l)\hspace{2mm}, \hspace{2mm}q_{(1+p)/2}(\hat{\mu}_n^u,\hat{k}_n^l)\Big)\label{option five}
\end{eqnarray}
where $\hat{k}_n^l$ is the lower limit of a two-sided $100(1-\alpha)\%$ confidence interval for $k$.  Here F$(y;\mu,k)$ has been parameterized such that the variance of $Y$ is a decreasing function of $k$, hence the use of $\hat{k}_n^l$.  In situations where the exact form of F$_{\scaleto{\sum Y}{5pt}}$ is not known an approximation can be used. 
\\

Estimated standard errors for Equations (\ref{option one}) and (\ref{option four}) are easily obtained using standard output from statistical packages such as Proc Genmod or Proc Glimmix, as are confidence limits for Equations (\ref{option two}) and (\ref{option five}). For Equation (\ref{option three}) when $N-n=1$, estimated standard errors can be easily produced using Proc LifeReg; when $N-n>1$, the delta method can be employed and solved numerically.  See SAS documentation for Proc QuantReg for semi-parametric inference on population percentiles when $N-n=1$.
\\

Wang and Wu\cite{wang2018} provide a summary of the literature on approximate tolerance and prediction limits for the gamma distribution\cite{bain1984, ashkar1998} and add their own approach to both. Similarly, Krishnamoorthy\cite{krishnamoorthy2011, krishnamoorthy2009} provides a summary and offers methods for the binomial and Poisson distributions. Lawless\cite{lawless2005} offers an approach using the probability integral transformation with Monte Carlo simulation to construct prediction intervals, and Shen\cite{shen2018} produces frequentist prediction intervals by mimicking Bayesian techniques. Equations (\ref{option one}) through (\ref{option five}) are comparatively easy to implement, particularly when adjusting for covariates in a regression setting, and based on their construction are applicable to a wide range of models. They are computationally more efficient than ``pure prediction'' machine learning algorithms based on the jackknife and techniques that rely on the bootstrap, Monte Carlo, or Markov Chain Monte Carlo, and can be continuously updated in a conformal manner.\cite{xie2020, vovk2019} 
\\

\section{Simulation}\label{simulation}

To verify the performance of the prediction limits considered above using a log link function and model-based standard error estimator we conducted a simulation experiment where a sample of $n$ observations are drawn from a Gamma$(k,\mu/k)$ distribution and used to predict the sum of the remaining $N-n$ observations.  
For comparison we have also investigated: i) a prediction interval using the  plug-in sampling distribution Gamma$\big((N-n)\hat{k},\hat{\mu}/\hat{k}\big)$, ii) an asymptotic prediction interval formed by pivoting the quantity $\sum_{i=n+1}^{N} Y_i/(N-n)\bar{Y}_n\sim F_{\scaleto{2(N-n)\hat{k},\hspace{0.5mm}2n\hat{k}}{7pt}}$ treating $\hat{k}$ as known, and iii) the approximate two-sided tolerance intervals using Equations (\ref{option three}), (\ref{option four}), and (\ref{option five}) for estimating the middle $100p\%$ of F$_{\scaleto{\sum Y}{5pt}}$ 
with $100(1-\alpha)\%$ confidence. Parameter estimates and their model-based standard errors were calculated using Proc Genmod.  Appendix \ref{simulation results} shows the coverage probabilities of these intervals over 10,000 simulations for various confidence levels, sample sizes, and parameter values. The coverage probabilities for Equations (\ref{option one}) and (\ref{option two}) are at or nearly equal to the nominal level for most sample sizes and clearly outperform the simple plug-in sampling distribution as a method for prediction in small to moderate sample sizes. The interval based on the $F$ distribution also performs close to the nominal level. Equation (\ref{option one}) is preferable over Equation (\ref{option two}) when $N-n>1$.  Equation (\ref{option two}) is preferable when $N-n=1$, particularly when $k\le 1$.  When $N=2n$ Equation (\ref{option two}) is most conservative as is expected from the discussion above in the context of a normal model. 
The approximate tolerance intervals in Equations (\ref{option three}), (\ref{option four}), and (\ref{option five}) perform as expected.  Equation (\ref{option three}) is preferable over Equations (\ref{option four}) and (\ref{option five}), except when $n$ is small and $N-n=1$.  Even when $N-n=1$ Equation (\ref{option four}) performs well; when $k\le1$ and $N-n=1$ Equation (\ref{option four}) is most conservative.  For Equation (\ref{option three}) simulation results are based on model-based output and the delta method.  
In the absence of an exact pivotal quantity one might be willing to accept the approximate prediction and tolerance intervals in Equations (\ref{option one}) through (\ref{option five}). 
\\

To investigate a different specification of the data generative process in the context of clinical trial recruitment we have considered a site-level Exponential$\big(\mu_j\big)$ interarrival time data process where the site-level $\lambda_j=1/\mu_j$ follows a Gamma$\big(\alpha,\beta\big)$ distribution over repeated trials, $j$ indexing site. This results in study-level interarrival times $Y_1,...,Y_n$ that follow an Exponential$\big(\mu=1/\sum_j\lambda_j\big)$ distribution within a single clinical trial where $\lambda_j$ is unknown and fixed, time to complete recruitment $\sum_{i=n+1}^N Y_i$ that follows a Gamma$\big((N-n), \mu\big)$ distribution within a single trial where $\lambda_j$ is unknown and fixed, and time to complete recruitment $\sum_{i=n+1}^N Y_i$ that follows a Pearson type VI distribution over repeated clinical trials where $\lambda_j$ is considered random. That is, 

\begin{quote}
Step 1: For a given simulated clinical trial, generate site-specific rate parameters $\lambda_j$ according to Gamma$\big(\alpha,\beta\big)$.
\\
Step 2: Holding each $\lambda_j$ fixed for a given clinical trial, generate times between next enrollments within each site using Exponential$\big(\mu_j\big)$, equivalent to sampling counts from Poisson$(\lambda_j)$.
\\
Step 3: Perform Steps 1 and 2 for each simulated clinical trial.
\end{quote}
This can be seen as the result of a Poisson-gamma process\cite{anisimov2007} where the observable site- and study-level recruitment rates follow Poisson$(\lambda_j)$ and Poisson$(\sum_j\lambda_j)$ distributions within a single trial where $\lambda_j$ is unknown and fixed, and the observable study-level recruitment rate follows a negative binomial distribution over repeated clinical trials where $\lambda_j$ is considered random.  
\\

The Pearson type VI distribution is a less tractable form of the $F$ distribution that depends on hyper parameters $\alpha$ and $\beta$.  Intervals from the Pearson type VI distribution could be used for making predictions about time to complete recruitment without forming a pivotal quantity.  The hyper parameters must either be assumed, estimated from the observed data and replaced using the plug-in principle, or ``updated'' using a Bayesian approach.  In this simulation scenario, however, we continue to simply model the study-level interarrival time in days as $Y_i\sim$ Gamma$\big(1,\mu\big)$, i.e. Exponential$\big(\mu\big)$, for $i=1,...,n$ with $\mu$ fixed and unknown so that the remaining time to complete recruitment $\sum_{i=n+1}^N Y_i$ follows a Gamma$\big((N-n), \mu\big)$ distribution, without any regard for random parameters. Because the scale parameter is held at $k=1$ the interval based on the $F$ pivotal quantity is most appropriate, though for comparison we have examined Equations (\ref{option one}) and (\ref{option two}) and the simple plug-in interval. As expected, Equations (\ref{option one}) and (\ref{option two}) and the prediction interval based on the $F$ ancillary pivotal quantity maintain the nominal coverage probability despite the model parameters changing from one simulated run to the next. This holds in general regardless of the distribution for $\lambda_j$. Therefore, the idea that the true underlying site- and study-level parameters randomly change from one repeated experiment to the next, and that the distribution of the parameters is somehow known or dependent on the observed data, is an unnecessary and unverifiable modeling assumption. 
\\

In reality the site- and study-level parameters are not random variables, and so the above Poisson-gamma framework is incorrect.  It does, however, demonstrate that even if site- and study-level parameters are somehow random variables there is no need to consider them as random.  A more appropriate simulation is to hold the unknown site- and study-level parameters fixed from one simulated clinical trial to the next and investigate the coverage probability of the prediction intervals.  This reflects the fact that the site- and study-level parameters, although unknown, are not sampled from a probability distribution and would not be different if the clinical trial recruitment was indeed repeated.   

\section{Application to Clinical Trial Recruitment}\label{application}
\subsection{Predicting Time to Full Recruitment}\label{application gamma}
\begin{figure}[H]
\centering
\includegraphics[trim={0.9cm 17.9cm 0 0}, clip, height = 2.3in]{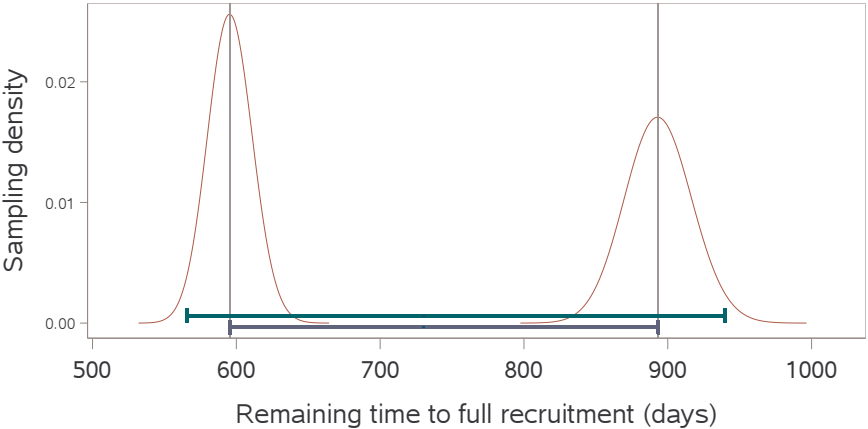}
\caption{\small{Estimated sampling distributions for $\sum_{i=n+1}^{N} Y_i$, the remaining time to complete recruitment, based on the lower and upper 95\% confidence limits of $(N-n)\mu$. Approximate 95\% prediction interval is formed using the $2.5^{th}$ and $97.5^{th}$ percentiles of these estimated distributions. See Equation (\ref{option two}). }} 
\label{sampling}
\end{figure}

\begin{figure}[H]
\centering
\includegraphics[trim={0.9cm 17.9cm 0 0}, clip, height = 2.3in]{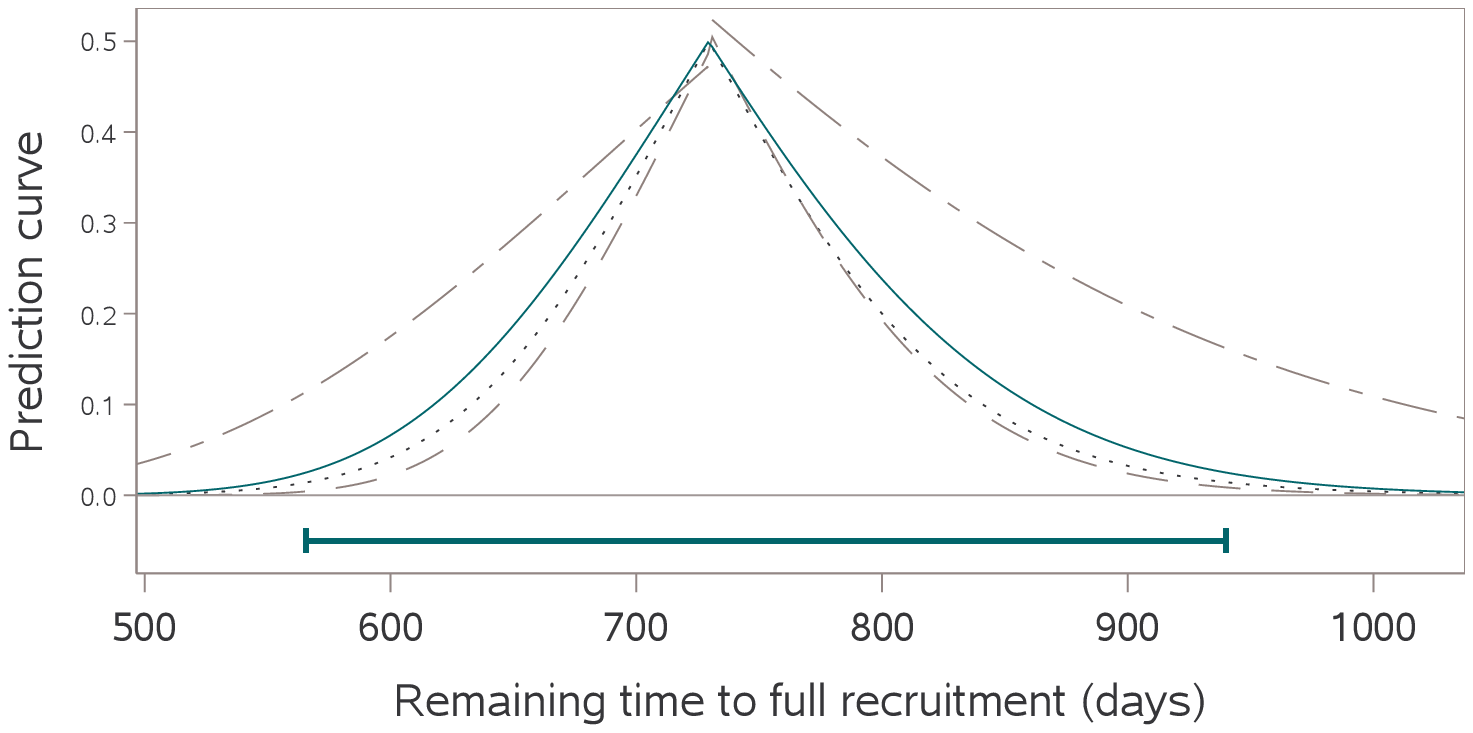}
\caption{\small{Solid curve depicts prediction intervals of all levels for $\sum_{i=n+1}^{N} Y_i$, the remaining time to full recruitment, using Equation (\ref{option two}). Error bars indicate the 95\% prediction limits. Dotted curve shows prediction intervals corresponding to Equation (\ref{option one}). Dashed curve shows prediction intervals corresponding to the pivotal quantity $\sum_{i=n+1}^{N} Y_i/(N-n)\bar{Y}_n\sim F_{\scaleto{2(N-n)\hat{k},\hspace{0.5mm}2n\hat{k}}{7pt}}$ 
treating $\hat{k}$ as known. Long-short curve shows prediction intervals using the $F$ pivotal quantity while holding $k=1$.}} 
\label{prediction}
\end{figure}

We consider the setting where $Y_1,...,Y_n,...,Y_N$ 
are the independent study-level interarrival times in days for $N$ total subjects recruited into a clinical trial. Based on $n$ currently enrolled subjects, our aim is to predict the remaining time to complete recruitment, $\sum_{i=n+1}^{N} y_i$. 
We demonstrate the intuitive approaches investigated above on a single simulated data set of $n=20$ interarrival times, shown below in Figure \ref{histogram}. Using a Gamma$\big(k,\mu/k\big)$ model for the interarrival times $Y_1,...,Y_{20}$ the maximum likelihood 
point estimate for $\mu$ is 2.61 days and the 95\% confidence limits are (2.13, 3.19) based on a Wald test using the robust sandwich standard error estimate and a log link function. The point and interval estimates for $k$ are 5.22 (2.69, 9.03). 
The goal is to predict the remaining time to complete recruitment of $N-n=300-20$ subjects. It follows that $\sum_{i=n+1}^N Y_i \sim$ Gamma$\big((N-n)k,\mu/k\big)$ with expectation $(N-n)\mu$.  Multiplying the point and interval estimates for $\mu$ by $N-n$ yields inference for the mean time to complete recruitment. We are 95\% confident $(N-n)\mu$ is between 596 and 893 days. Using Equation (\ref{option two}), if this mean is truly 596 we would observe a time to complete recruitment less than 566 an estimated 2.5\% of the time, and if this mean is truly 893 we would observe a time to complete recruitment greater than 940 an estimated 2.5\% of the time. This is seen in Figure \ref{sampling}. In this way we are 95\% confident the remaining time to complete recruitment will be between 566 and 940 days. 
Using Equation (\ref{option one}) with a log link function and robust sandwich standard error estimate, the one-sided p-values testing the hypotheses that the remaining time to complete recruitment will be $\le$ 583 or $\ge$ 914 are both $2.5\%$. Therefore we are $95\%$ confident the remaining time to complete recruitment will be between 583 and 914. The prediction confidence curves depicted in Figure \ref{prediction} show p-values and prediction intervals of all levels for hypotheses about the remaining time to complete recruitment.\cite{johnson2021, lawless2005, shen2018} Based on our simulation results, the dotted curve corresponding to Equation (\ref{option one}) is most appropriate.  SAS code is provided in Appendix \ref{sas code} for producing the figures below.  See Appendix \ref{additional figures} for additional figures depicting prediction densities.
\\

\begin{figure}[H]
\centering
\includegraphics[trim={1.9cm 18cm 0 0}, clip, height = 2.2in]{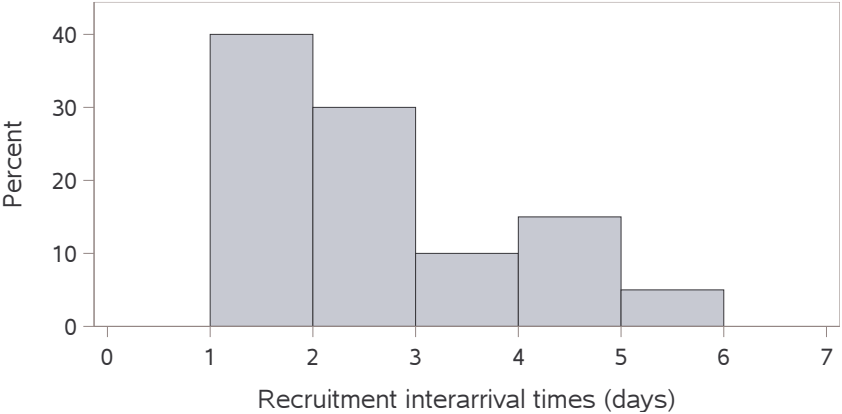}
\caption{\small{Histogram of the $n=20$ interarrival times. }} 
\label{histogram}
\end{figure}

Figure \ref{prediction} also shows prediction intervals based on the $F$ distributed pivotal quantity.  
The Pearson type VI distribution from a Poisson-gamma model discussed in Section \ref{simulation} requires sophisticated modeling software, yet if implemented ``objectively'' manages to produce results essentially identical to the prediction limits based on the $F$ pivotal quantity, and similar to Equations (\ref{option one}) and (\ref{option two}), while holding $k=1$. This coincidence is due to the connection between frequentist and objective Bayesian methods.  
Recall that interest surrounds predicting recruitment for the sites at hand, not a broader population of sites.  Adopting an objective definition of probability, the correct sampling unit in this problem is not site, it is subject.  This is seen by consulting the literature on linear mixed models.  Therefore, sites should not be treated as random.  These limits $\big([N-n]\bar{y}_n\cdot f_{\scaleto{\alpha/2,\hspace{0.5mm}2(N-n),\hspace{0.5mm} 2n}{7pt}}\hspace{1mm},$ $\hspace{2mm}[N-n]\bar{y}_n\cdot f_{\scaleto{1-\alpha/2,\hspace{0.5mm}2(N-n),\hspace{0.5mm}2n}{7pt}}\big)$ based on the $F$ pivotal quantity and the corresponding prediction confidence curve 
\begin{eqnarray}
C\big(\textstyle\sum y\big)&\equiv& \left\{ \begin{array}{cc}
P\Big(F_{\scaleto{2n,\hspace{0.5mm}2(N-n)}{7pt}}>\frac{(N-n)\bar{y}_n}{\sum y}\Big) & \text{if } \sum y < (N-n)\bar{y}_n\\
 &  \nonumber\\
 P\Big(F_{\scaleto{2n,\hspace{0.5mm}2(N-n)}{7pt}}<\frac{(N-n)\bar{y}_n}{\sum y}\Big)  & \text{if }\sum y > (N-n)\bar{y}_n \end{array}  \right.\nonumber
\end{eqnarray}
are easily programmed in a spreadsheet application such as Microsoft Excel and can be provided to a practitioner with limited knowledge of statistics to be updated using the observed data. $\bar{y}_n$ is calculated as the number of days it took to recruit $n$ subjects divided by $n$, and $f_{\scaleto{\alpha/2,\hspace{0.5mm}2(N-n),\hspace{0.5mm} 2n}{7pt}}$ is the $100(\alpha/2)^{th}$ percentile from the $F$-distribution with numerator and denominator degrees of freedom $2(N-n)$ and $2n$ respectively. 
If sophisticated modeling is considered in practice, the $\sum_{i=n+1}^N Y_i \sim$ Gamma$\big((N-n)k,\mu/k\big)$ model presented at the beginning of Section \ref{application gamma} has a free scale parameter making it more flexible in modeling the study-level interarrival time than the Gamma$\big((N-n),\mu\big)$ distribution contained within the Poisson and Poisson-gamma processes.  Forming prediction intervals without imposing distributional assumptions on model parameters reduces overall model complexity compared to the Poisson-gamma process while maintaining the nominal coverage probability.  Furthermore, the prediction limits in Equation (\ref{option one}) are applicable to parametric models other than the gamma distribution, as well as semi- and non-parametric models, and can be constructed while incorporating historical data. 
\\

Some authors refer to the Pearson type VI distribution discussed above as a `within-trial' model presumably since $n$ observations from the current trial are used to predict the remaining recruitment time of the trial at hand,\cite{bakhshi2013} but such a distinction might suggest $1/\mu_j=\lambda_j$ is in flux from one recruited subject to the next, from one infinitesimal moment to the next within a single clinical trial.  The same authors contrast this with a `between-trial' model, requiring further distributional assumptions on the hyper parameters $\alpha$ and $\beta$.  `Between-trial' could be taken to mean 
the data collected are heterogeneous from different programs, and the distributions on $\alpha$ and $\beta$ are meant to reflect this heterogeneity when predicting the recruitment of a single clinical trial.  
Using the approaches explored in Equations (\ref{option one}) and (\ref{option two}) for prediction, this program heterogeneity 
would simply be captured in the standard error of $\hat{\mu}$.  Alternatively, where possible one might naturally adjust for or stratify by program to reduce the standard error and make the predictions program-specific.

\subsection{Predicting Number of Subjects Recruited}\label{modeling counts}
\begin{figure}[H]
\centering
\includegraphics[trim={0.9cm 17.9cm 0 0}, clip, height = 2.3in]{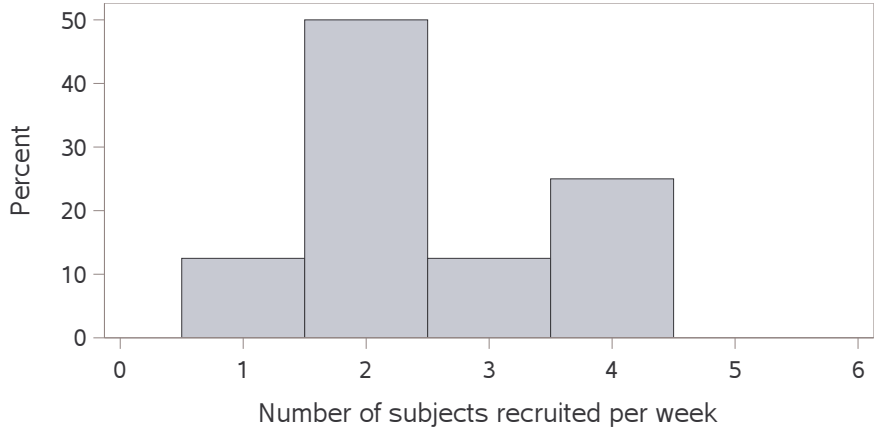}
\caption{\small{Histogram showing the distribution of number of subjects recruited per week based on $n=20$ subjects. }} 
\label{histogram counts}
\end{figure}

\begin{figure}[H]
\centering
\includegraphics[trim={0.9cm 17.9cm 0 0}, clip, height = 2.3in]{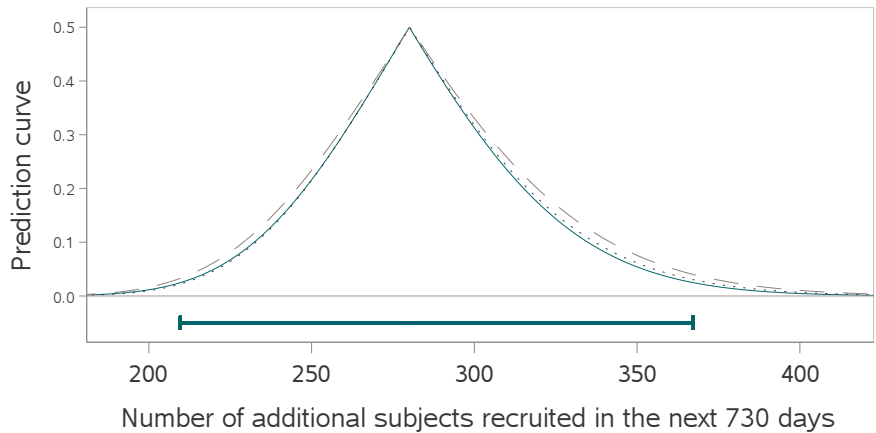}
\caption{\small{Solid curve shows prediction intervals of all levels for the number of additional subjects recruited in the next 730 days (2 years) using Equation (\ref{option two}). Error bars indicate the 95\% prediction limits. Dotted curve shows prediction intervals corresponding to Equation (\ref{option one}). Dashed curve shows prediction intervals using the method of Krishnamoorthy and Peng (2011) while incorporating the dispersion parameter.}} 
\label{prediction counts}
\end{figure}

Let $X_1,...,X_d,...,X_D$ be the independent study-level number of subjects recruited per day in a clinical trial.  Based on $d$ observations, our aim is to predict the number of additional subjects recruited in the next $D-d$ days, $\sum_{l=d+1}^D x_l$.  For the same $n=20$ subjects investigated previously, the number of subjects recruited per week is shown in Figure \ref{histogram counts}. Based on the currently enrolled subjects, our aim is to predict the number of additional subjects recruited in the next 730 days. We consider a Poisson$(\phi,\lambda)$ model for the study-level number of subjects recruited per day with a free dispersion parameter $\phi$.  The maximum likelihood and 95\% confidence interval estimates for the exposure adjusted mean number of subjects recruited per week are 2.69 (2.20, 3.28). The dispersion scale estimate is 0.46. Multiplying the point and interval estimates for the mean recruitment rate by 730/7 yields inference for the mean number of subjects recruited per 730 days. We are 95\% confident the mean number of subjects recruited per 730 days is between 229 and 342. Using Equation (\ref{option two}), if this mean is truly 229 we would observe less than 210 additional subjects recruited an estimated 2.5\% of the time, and if this mean is truly 342 we would observe more than 367 additional subjects recruited an estimated 2.5\% of the time. In this way we are 95\% confident the number of additional subjects recruited in the next 730 days will be between 210 and 367. A gamma approximation to the Poisson sampling distribution was used to incorporate the dispersion parameter when forming these prediction limits.  Using Equation (\ref{option one}) with a log link function and model-based standard error estimate, the one-sided p-values testing the hypotheses that the number of additional subjects recruited will be $\le$ 211 or $\ge$ 372 are both $2.5\%$. Therefore we are $95\%$ confident the number of additional subjects recruited in the next 730 days will be between 211 and 372. Intervals constructed in either manner will cover the as of yet unobserved number of additional subjects recruited approximately 95\% of the time. Constructing prediction intervals of all levels produces the prediction confidence curves depicted in Figure \ref{prediction counts}.  This figure also shows prediction intervals using the method of Krishnamoorthy and Peng (2011) while incorporating the dispersion parameter.  
SAS code is provided in Appendix \ref{sas code}.  If a Poisson model is implemented without a free dispersion parameter, then the prediction intervals and prediction confidence curves corresponding to Equations (\ref{option one}) and (\ref{option two}) can be easily programmed in a spreadsheet application using the model-based standard error estimate.  
\\

Some authors who specialize in queueing and recruitment processes insist on utilizing a single, all-encompassing model to capture both interarrival time and number of subjects recruited, i.e. a Poisson process.  The argument is that such a process is memoryless and jointly models the endpoints of interest.  While this is certainly a noble pursuit and an interesting property for a model to have, it is by no means a requirement.  Our approach to separately model interarrival time and number of subjects recruited provides flexibility in model selection and highlights the principle of division of labor\cite{efron1986} allowing one to avoid construction of an all-encompassing model.  A Poisson process represents a special case of the models we have considered above by holding $k=1$ in the gamma model and $\phi=1$ in the Poisson model, and can certainly be utilized if it is a good fit to the data.   


\subsection{Time-varying Rate}\label{time-varying}
In the preceding applications we assumed a constant study-level mean recruitment rate, suggesting all sites begin recruiting at the same time point. These examples may seem simple, but they were chosen on purpose to make the prediction intervals comparable to the models explored by Anisimov and Federov\cite{anisimov2007} and others,\cite{gajewski2008} and to demonstrate how simply the recruitment problem can be addressed.  In contrast, it may naturally arise that part way through a study additional sites are opened, existing sites are closed, there is a push in recruitment activity, or there are breaks in recruitment for weekends, holidays, etc.  Several authors propose complex heterogeneous site-level models with random parameters to account for variability in the observable study-level recruitment rate and longitudinal change in the mean.\cite{anisimov2011, lan2019}  However, heterogeneity across sites is equally well captured in the standard error of the estimated study-level mean.  
Additionally, even with staggered site initiation it may be reasonable to assume a constant study-level mean rate over time.  
This has been noted elsewhere in the literature on recruitment.\cite{minois2017, gajewski2012, lan2019}  The REMATCH study published by Lan et al. (2019) is an excellent example of a near constant study-level mean recruitment rate despite fluctuations in the number of active sites.  
Nevertheless, if an irrefutable trend in the recruitment rate is observed it is important to capture this in order to make accurate predictions.  Next we consider a site-level approach for addressing staggered site initiation, followed by a generalized linear model to account for variability in the observable study-level rate and a longitudinal trend in the mean.
\\

Let $l=1,2,...,d,...,D$ be an index for the $l^{th}$ day of recruitment, and let $X_{lj}$ be the observable number of subjects recruited in the $j^{th}$ site on the $l^{th}$ day, with fixed expectation $\lambda_j$.  Let $s_l$ be the number of actively recruiting sites on the $l^{th}$ day.  This collection of sites \textit{is} the distribution of sites of interest, not a sample from a broader distribution of sites.  This fixed collection of sites induces a mixture distribution of $X_{lj}$ over $j$.  The sample mean of this mixture distribution is $\frac{1}{s_l}\sum_{j=1}^{s_l}X_{lj}$ with expectation $\frac{1}{s_l}\sum_{j=1}^{s_l}\lambda_j$, the average recruitment per site on the $l^{th}$ day.  Some sites may be over performers, others may be under performers, and on average 
a site will recruit $\frac{1}{s_l}\sum_{j=1}^{s_l}\lambda_j$ subjects 
in a given day.  Then $W=\sum_{l=d+1}^{D}\sum_{j=1}^{s_l}X_{lj}$  is the observable number of additional subjects recruited in the remaining $D-d$ days with expectation $\sum_{l=d+1}^D\sum_{j=1}^{s_l}\lambda_j$.    
If the mixture distribution of $X_{lj}$ over $j=1,...,s_l$ is the same or approximately the same for any $l^{th}$ day, then one can view $\sum_{l=1}^d \sum_{j=1}^{s_l}X_{lj}/\sum_{l=1}^d s_l$ as an estimator for $W/\sum_{l=d+1}^{D}s_l$ and both of these as estimators for a commonly defined parameter 
$\theta\equiv\frac{1}{s_l}\sum_{j=1}^{s_l}\lambda_j$, the mean number of subjects recruited per site per day.  The pivotal quantity corresponding to Equation (\ref{option one}) becomes 
\begin{eqnarray}\label{site_level_prediction}
\big[g\{\hat{\theta}_{d^*}\}-g\{W/(D^*-d^*)\}]/\hat{\text{SE}}_{d^*}
\overset{asymp}{\sim}N(0,1)
\end{eqnarray}
where $D^*=\sum_{l=1}^{D}s_l$ is the number of site-days up to day $D$, $d^*=\sum_{l=1}^{d}s_l$ is the number of site-days up to day $d$, $\hat{\theta}_{d^*}=\sum_{l=1}^d \sum_{j=1}^{s_l}X_{lj}/\sum_{l=1}^d s_l$ is an estimator for $\theta$ using the number of subjects recruited over the first $d^*$ site-days, and $\hat{\text{SE}}_{d^*}$ is a model-based or empirical estimator for the standard error of $\big[g\{\hat{\theta}_{d^*}\}-g\{W/(D^*-d^*)\}]$.  
Inverting a hypothesis test for $W$ allows one to identify a range of plausible values while acknowledging that $\theta$ and the constituent site-specific rates are unknown fixed quantities.  The estimate for $g\{\theta\}$ and associated standard error can be calculated in Proc Genmod or a similar procedure by modeling $X_{lj}$ $l=1,...,d$, $j=1,...,s_l$ with an intercept only.  
The model definition for $\theta$ above was chosen in order to make $W$ identifiable from the observed data.  Other definitions could be entertained and the pivotal quantity adjusted accordingly.  
For instance, one might instead estimate $\theta$ using $X_{dj}$ $j=1,...,s_d$ 
and assume that $\theta$ is the same for any day $l=d,...,D$.  Although the model is satisfied when $\lambda_j\equiv\lambda$ $\forall j$, this is not a necessary condition.  
\\

The approach above requires a schedule of site openings and closures over the remaining $D-d$ days.  It allows for heterogeneity across sites while assuming that the average recruitment rate per site is constant over time, and attributes any change in the study-level mean recruitment rate to the number of active sites.  It is useful for investigating the effect of opening additional sites or closing existing sites when predicting study-level recruitment.  Another solution is to extend the approach from Section \ref{modeling counts} to model the study-level mean recruitment rate as a function of time, $\lambda(l)$, using a generalized linear model (linear in the coefficients).  
Proc Genmod or Proc NLMixed can be used to explore link functions, constraints on parameter estimates, polynomial regression terms, and piecewise regression models.   A log- or root-transformed regressor or a power link function would allow the modeled mean number of subjects recruited per day (or week or month) to increase at a decreasing rate over time.  This would account for a changing study-level mean rate for any number of reasons, not simply the number of active sites.  Predictions for the number of subjects recruited on the $l^{th}$ day (or week or month) would be made using Equations (\ref{option one}) or (\ref{option two}), i.e.
\begin{eqnarray}\label{daily rate}
\big[g\{\hat{\lambda}_d(l)\}-g\{ x_l\}\big]\big/\hat{\text{SE}}_d
\overset{asymp}{\sim}N(0,1)
\end{eqnarray}
where $\hat{\lambda}_d(l)$ is an estimate of the mean recruitment rate on the $l^{th}$ day using information up to day $d$, and $\hat{\text{SE}}_d$ is a model-based or empirical estimator for the standard error of $\big[g\{\hat{\lambda}_d(l)\}-g\{ x_l\}\big]$.  
Furthermore, the modeled mean could be extrapolated and summed over the next $D-d$ days to estimate the additional number of subjects recruited, with predictions made using Equations (\ref{option one}) or (\ref{option two}), i.e.
\begin{eqnarray}\label{sum rate}
\big[g\{\textstyle\sum_{l=d+1}^D\hat{\lambda}_d(l)\}-g\{\sum_{l=d+1}^D x_l\}\big]\big/\hat{\text{SE}}_d
\overset{asymp}{\sim}N(0,1)
\end{eqnarray}
where $\hat{\text{SE}}_d$ is a model-based or empirical estimator for the standard error of $\big[g\{\textstyle\sum_{l=d+1}^D\hat{\lambda}_d(l)\}-g\{\sum_{l=d+1}^D x_l\}\big]$.  
Inverting a hypothesis test for $\sum_{l=d+1}^D x_l$ allows one to identify a range of plausible values for the number of subjects recruited while acknowledging that $\lambda(l)$ and the constituent site-specific rates are unknown fixed quantities.  Inverting this quantity for $D$ allows one to identify the remaining number of days (or weeks or months) of recruitment, $D-d$, for which it is estimated the recruitment target will be met.  Similarly, one can also identify the range of $D-d$ over which the $100(1-\alpha)\%$ prediction interval for the number of subjects recruited contains the recruitment target.  
This simple extension using a generalized linear model can also be applied to model the study-level mean interarrival time as a function of the $i^{th}$ subject recruited, $\mu(i)$.   
An exponentiated or inverse root-transformed regressor or an inverse power link function would account for the study-level mean interarrival time decreasing at a decreasing rate as more subjects are enrolled, be it for staggered site initiation, a push in recruitment, or any other reason.  Predictions about the remaining time to complete recruitment could be made using Equations (\ref{option one}) or (\ref{option two}), i.e.
\begin{eqnarray}\label{sum_interarrival}
\big[g\{\textstyle\sum_{i=n+1}^N\hat{\mu}_n(i)\}-g\{\sum_{i=n+1}^N y_i\}\big]\big/\hat{\text{SE}}_n
\overset{asymp}{\sim}T_{n-1}
\end{eqnarray}
where $\hat{\mu}_n(i)$ is an estimate of the mean interarrival time when the $i^{th}$ subject is recruited using information from $n$ subjects, and $\hat{\text{SE}}_n$ is a standard error estimator for $\big[g\{\textstyle\sum_{i=n+1}^N\hat{\mu}_n(i)\}-g\{\sum_{i=n+1}^N y_i\}\big]$.  
\\

\begin{figure}[H]
\centering
\includegraphics[trim={1.2cm 17.9cm 0 0}, clip, height = 2.3in]{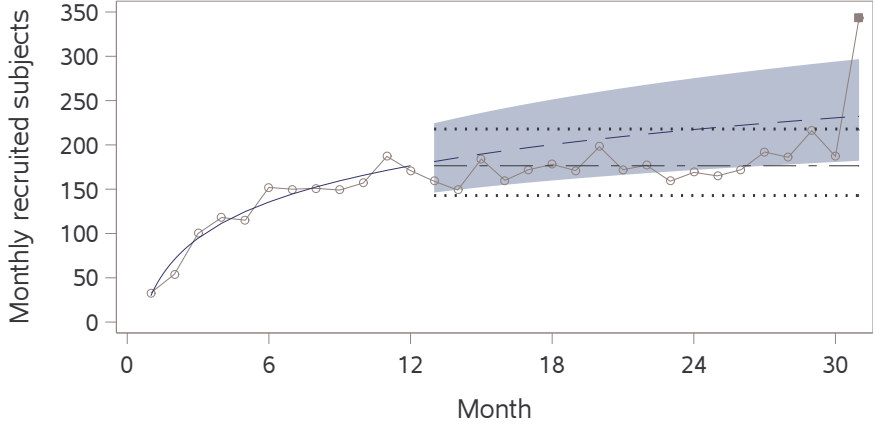}
\caption{\small{Observed study-level monthly recruitment in gray circles connected as a series.  Mean model for the number of subjects recruited per month in solid blue.  Nonlinear model extrapolation in dashed blue curve with 95\% prediction limits as a solid band.  Linear extrapolation in long-short black curve with dotted 95\% prediction limits. }} 
\label{b_38}
\end{figure}
\begin{figure}[H]
\centering
\includegraphics[trim={0.9cm 17.9cm 0 0}, clip, height = 2.3in]{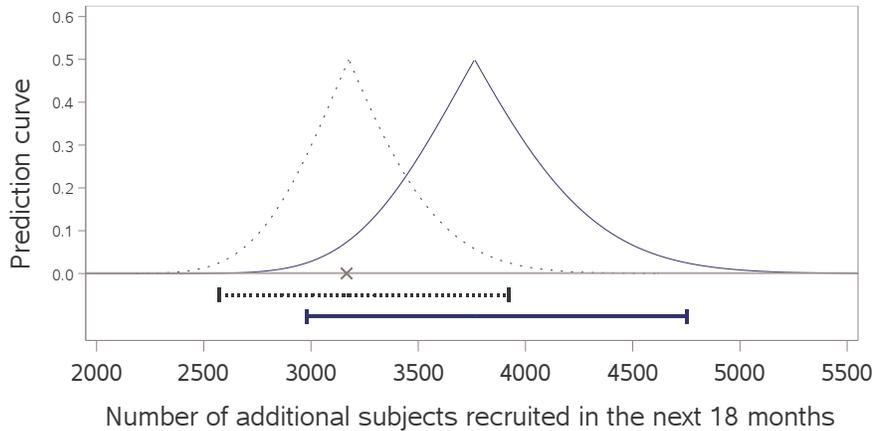}
\caption{\small{Solid curve shows prediction intervals of all levels for the number of additional subjects recruited in months 13 through 30 using Equation (\ref{sum rate}) by summing the nonlinear extrapolation of the modeled monthly rate.  Dotted curve shows prediction intervals of all levels using Equation (\ref{sum rate}) by summing the linear extrapolation of the modeled monthly rate.  Error bars indicate 95\% prediction limits.  Gray x indicates the observed number of subjects recruited.    }} 
\label{b_38_sum}
\end{figure}

Figure \ref{b_38} shows the 
observed recruitment for the National Surgical Adjuvant Breast and Bowel Project (NSABP) B-38 study published by Lan et al. (2019).\cite{lan2019}  The observed study-level monthly recruitment is plotted in gray circles connected as a series.  The last data point in the series represents a burst of activity in the final days of recruitment, prorated for an entire month.  The solid blue curve shows the study-level mean model for the number of subjects recruited per month over the first 12 months fit in Proc Genmod using a Poisson model with a free dispersion parameter, 
a log transformation of month as the regressor, and an identity link function.  Similar results are obtained using a $20^{th}$ root or higher-order regression term for month.  The blue dashed curve shows the model extrapolation over the next 19 months, and the solid band identifies the 95\% prediction limits based on Equation (\ref{daily rate}) in terms of months with a log link function.  
The black long-short curve shows a linear extrapolation assuming a constant recruitment rate using the last fitted regression value, with dotted 95\% prediction limits formed using Equation (\ref{daily rate}) and the delta method with a log link function.  Both extrapolation methods do well at predicting the observed monthly recruitment rate, though the non-linear prediction fails to cover a few observations.  Figure \ref{b_38_sum} shows p-values and prediction intervals of all levels for the number of additional subjects recruited for months 13 through 30.  Nonlinear extrapolation of the modeled monthly rate was summed over months 13 through 30 to produce the solid blue prediction confidence curve using Equation (\ref{sum rate}) and the delta method with a log link function.  
Linear extrapolation of the modeled monthly rate was summed over months 13 through 30 to produce the dotted black prediction confidence curve using Equation (\ref{sum rate}) and the delta method with a log link function.  The observed number of additional subjects recruited in this 18 month period 
is indicated by the gray x.  Both the linear and non-linear extrapolations do well at predicting the observed number of subjects recruited.  In a similar fashion the study-level mean interarrival time can be modeled, extrapolated, and summed to form predictions for the remaining time to complete recruitment using Equation (\ref{sum_interarrival}); however, Lan et al. (2019) only published aggregate monthly study-level recruitment data for B-38 seen above in Figure \ref{b_38}.  To implement Equation (\ref{site_level_prediction}) ideally one would use site-level daily recruitment data since the number of active sites may not remain constant for an entire month.   

If site-, region-, or country-level recruitment predictions are required the prediction intervals explored here can be applied to a stratified or adjusted model.  If there is an attrition rate competing against the enrollment rate, the methods explored are equally applicable for modeling and predicting attrition.  A pivotal quantity for enrollment can be combined with a pivotal quantity for attrition to predict the number of actively enrolled subjects.  If uncertainty surrounds the number of initiated or active sites, these methods are also equally applicable.  To forecast recruitment when little or no observed data are available, the approaches we have considered can incorporate historical or hypothetical data assumed to be exchangeable with the current recruitment process.  A prediction other than model extrapolation can be considered by simply positing a model and standard error, analogous to a pattern mixture model in a missing data problem.
\\

A longitudinal or time series approach could be further extended to account for shocks to the recruitment process.  For instance, consider a world-wide event that affects the recruitment rate for all sites within a study. Suppose that predictions obtained elsewhere indicate the remaining effects of this event will last 1095 days (36 months). The data collected so far after the event took place can be used to construct a pivotal quantity for the total number of subjects recruited in the next 1095 days.  A similar pivotal quantity for after the event is over can be constructed by assuming the recruitment rate returns to its previous levels, using information obtained before the world-wide event.  These independent pivotal quantities can be combined to form a single pivotal quantity for the additional number of subjects recruited.  Uncertainty around the 1095 day prediction can also be incorporated.

\section{Application to Real-world Time on Treatment}\label{tot}
\begin{figure}[H]
\centering
\includegraphics[trim={1.4cm 17.9cm 0 0}, clip, height = 2.3in]{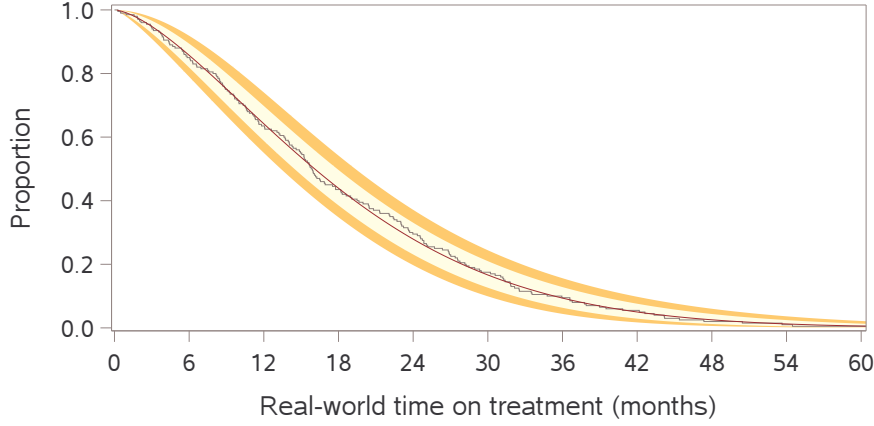}
\caption{\small{Step function shows Kaplan-Meier estimate of time on treatment survival curve.  Solid curve shows fitted Weibull model.  Yellow bands indicate 95\% tolerance limits for population percentiles.  Dark yellow bands indicate 95\% prediction limits for an observable percentile estimate in a repeated experiment with 100 events. }} 
\label{tot_survival}
\end{figure}

\begin{figure}[H]
\centering
\includegraphics[trim={-0.1cm 17.9cm 0 0}, clip, height = 2.3in]{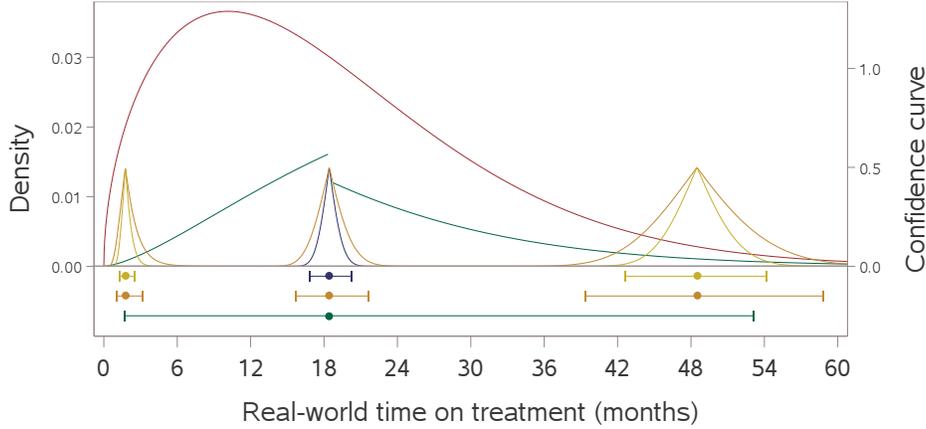}
\caption{\small{Density depicts estimated Weibull distribution of time on treatment.  Blue confidence curve shows inference on the population mean.  Yellow confidence curves show inference on the $2.5^{th}$ and $97.5^{th}$ population percentiles.  Dark yellow confidence curves show predictive inference on the observable mean and percentiles in a repeated experiment with 100 events.  Green confidence curve depicts a subject-level prediction.  Corresponding error bars indicate 95\% intervals.}} 
\label{tot_density}
\end{figure}

 It is often of interest for a pharmaceutical company to characterize time on treatment, also known as time to treatment discontinuation, in a target patient population for a particular disease indication in a real-world setting.  This time on treatment data is collected via electronic health records and can be updated with additional observations on a weekly or monthly basis.    Let $Y_1,...,Y_n,...,Y_N$ be the observable subject-level time on treatment, defined as the time to treatment discontinuation, loss to follow-up, or death, whichever occurs first, with non-informative administrative censoring.  Using $n$ observations the goal is to estimate and infer population-level parameters of time on treatment in the target patient population, and to predict future experimental results based on $N-n$ observations.  Figure \ref{tot_survival} shows the results of an example constructed to mimic real-world time on treatment for an oncology indication.  The Kaplan-Meier estimate of the time on treatment survival function is plotted for a sample of $n=200$ subjects, along with the estimated Weibull curve using Proc Lifereg.  The light yellow band indicates 95\% tolerance limits for population percentiles using the pivotal quantity corresponding to Equation (\ref{option three}) with a log link function, $\big[g\{q_p\big(\hat{\mu}_n,\hat{k}_n\big)\}-g\{q_p\}\big]\big/\hat{\text{SE}}_{n,p}\overset{asymp}{\sim}T_{n-1}$, where in this example $q_p\big({\mu},{k}\big)$ is the quantile function for F$_{\scaleto{Y}{5pt}}(y;\mu,k)$, not F$_{\scaleto{\sum Y}{5pt}}(y;\mu,k)$.  The dark yellow bands indicate 95\% prediction limits for an observable percentile estimate in a repeated experiment with $N-n=100$ events using the pivotal quantity $\big[g\{q_p\big(\hat{\mu}_n,\hat{k}_n\big)\}-g\{q_p\big(\hat{\mu}_{N-n},\hat{k}_{N-n}\big)\}\big]\big/\sqrt{n}\cdot\hat{\text{SE}}_{n,p}\sqrt{1/n+1/(N-n)}\overset{asymp}{\sim}T_{n-1}$ with a log link function where $\hat{\text{SE}}_{n,p}$ is an estimator for the standard error of $g\{q_p\big(\hat{\mu}_n,\hat{k}_n\big)\}$.  Figure \ref{tot_density} shows the estimated Weibull density for time on treatment with inference on the population mean in blue, and inference on the population $2.5^{th}$ and $97.5^{th}$ percentiles in yellow.  Corresponding predictions are in dark yellow for a repeated experiment with $N-n=100$ events, and a subject-level prediction is in green for a single event.  The subject-level prediction was formed using Equation (\ref{option two}) with the quantile function for F$_{\scaleto{Y}{5pt}}(y;\mu,k)$.  The visualizations in Figures \ref{tot_survival} and \ref{tot_density} allow stakeholders to easily identify point and interval estimates for population-level quantities, as well as point and interval estimates for repeated experimental results.

\section {Application to Clinical Trial Success}\label{application success}
Here we consider the setting where $Y_1,...,Y_n$ are binary measurements indicating improvement in a disease activity score collected on $n$ clinical trial subjects in a small phase 2 study. Interest surrounds using the estimated odds ratio between treatment and control to predict the result of a larger phase 3 study on the same endpoint involving $m$ subjects from the same patient population. We demonstrate the approach investigated above on a single simulated data set of $n=100$ subjects. Using a binomial model for $Y_1,...,Y_n$ the maximum likelihood estimate for the odds ratio is $\hat{\rho}_{n}=3.75$, and the 95\% confidence limits are (1.03, 14.05) based on a logit link and model-based standard error estimate. The goal is to predict the observed odds ratio in phase 3 among $m=600$ subjects. 
If the treatment allocation ratio is the same or nearly the same in phase 2 and phase 3, the pivotal quantity corresponding to Equation (\ref{option one}) can be implemented as $\big[\text{log}\{\hat{\rho}(\boldsymbol{Y}_{n})\}-\text{log}\{\hat{\rho}(\boldsymbol{Y}_{m})\}\big]\big/\sqrt{n}\cdot\hat{\text{SE}}_{n}\sqrt{1/n+1/m}\overset{asymp}{\sim}T_{n-1}$ where $\hat{\text{SE}}_{n}$ is an estimator for the standard error of $\text{log}\{\hat{\rho}(\boldsymbol{Y}_{n})\}$.  
The one-sided p-values testing $H_0$: $\hat{\rho}_m\le 0.90$ and $H_0$: $\hat{\rho}_m \ge 15.62$ are both $2.5\%$. Therefore we are $95\%$ confident the observed odds ratio in phase 3 will be between 0.90 and 15.62, regardless of the unknown fixed true population-level odds ratio $\rho$.  
One can form a prediction interval for the observable phase 3 test statistic instead of the observable phase 3 treatment effect, though this has little impact on the prediction (see Figure \ref{prediction or z} in Appendix \ref{additional figures}). 
\\

\begin{figure}[H]
\centering
\includegraphics[trim={0.9cm 18cm 0 0}, clip, height = 2.2in]{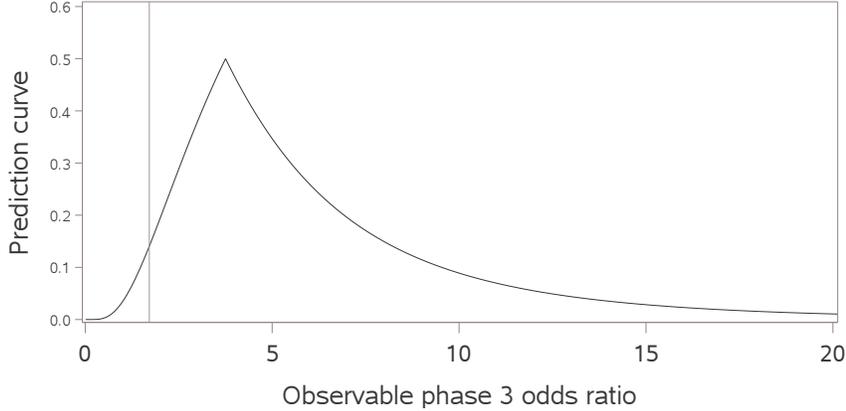}
\caption{\small{Solid curve shows p-values and prediction intervals corresponding to Equation (\ref{option one}).  The peak of the curve corresponds to the observed odds ratio in phase 2 based on $n=100$.  The reference line identifies the phase 3 minimum detectable effect, 1.71, and the height of the prediction confidence curve displays the p-value testing the hypothesis $H_0$: $\hat{\rho}_m\le 1.71$ .} }
\label{prediction or}
\end{figure}

The peak in Figure \ref{prediction or} corresponds to the observed odds ratio in phase 2 based on $n=100$.  The vertical reference line identifies the minimum detectable effect in phase 3, and the height of the prediction confidence curve displays the p-value testing this hypothesis.  If the minimum detectable effect for the planned phase 3 study is 1.71, the most we can claim is that we are 86\% confident that the phase 3 study will achieve a successful result ($\text{p-value}=0.14$ testing $H_0$: $\hat{\rho}_m \le 1.71$).  This $86\%$ confidence level is analogous to the Bayesian quantity \textit{probability of success}, or \textit{assurance}. In the Bayesian framework probability itself is not objectively defined, so that the observed $\text{log}\{\hat{\rho}_n\}$ and its estimated standard error are viewed as the mean and standard deviation of a random population-level log$\{\rho\}$ resulting from a vague prior, 
and the prediction density is interpreted as a legitimate probability distribution for the observable phase 3 odds ratio.  This leads to the conclusion that the as of yet unobserved $\hat{\rho}_m$ will be greater than or equal to 1.71 with 86\% ``probability.''  Several authors have proposed \textit{probability of success} as a more appropriate alternative to frequentist power and espoused its use as part of net present value calculations.\cite{temple2021,o2005, trzaskoma2007, chuang2006}  The claim is that one must assume a particular population-level treatment effect is true in order to calculate power, while \textit{probability of success} exists unconditionally on the population-level treatment effect.\cite{best2018} Viewed as the confidence level of a prediction interval or a p-value, \textit{probability of success} is a misnomer as it is not a probability statement about the phase 3 result.  Adopting an objective definition of probability, the population-level odds ratio $\rho$ is not a random variable that depends on the observed phase 2 data, nor is the observable phase 3 result dependent on the observed phase 2 data. While there is certainly value in predicting the phase 3 result, the actual probability of achieving end-of-study success is power and it may be more appropriate to perform inference on power when making a Go/No-Go decision into phase 3 to account for the uncertainty around estimating the population-level treatment effect.\cite{johnson2021} 

\section{Closing Remarks}\label{closing remarks}
We have considered intuitive tolerance and prediction intervals that perform well even in small sample sizes, and based on their construction are applicable to a wide range of models.  Equation (\ref{option two}) is most notable for its simplicity and performance when predicting a subject-level observation from a skewed distribution.  
The prediction intervals we have explored avoid distributional assumptions about model parameters. This approach reduces overall model and computational complexity compared to other methods for prediction while producing intervals that maintain their nominal coverage probability.    

\section*{Acknowledgment}
We are grateful to NRG Oncology for sharing the 31 aggregate recruitment values from the B-38 study.  

\pagebreak

\bibliographystyle{WileyNJD-AMA}
\bibliography{bib}

\pagebreak
\appendix
\section{Simulation Results}\label{simulation results}

\begin{table}[H]
\begin{center} 
\begin{tabular}{l l l r r r r r }
Data Generative  & Interval & Nom & n=10 & n=20 & n=100 & n=290 & n=299 \\
Process& & Cov& N=11& N=300 &N=300 & N=300 & N=300\\
& & Prob& & & & \\
\cline{1-1} \cline{2-2} \cline{3-3} \cline{4-4} \cline{5-5} \cline{6-6} \cline{7-7} \cline{8-8}
~ & ~ & ~ & ~ & ~ & ~ & ~&~\\
$Y_i\sim$ Gamma$\big(4,\frac{2.5}{4}\big)$ & Equation (\ref{option one}) & 0.95 &0.920  &0.947  & 0.943&0.950 &0.927\\
  &   & 0.80  &0.770   &0.790   &0.797  &0.792  &0.778 \\
 &   &  0.50  &0.477   & 0.492  &0.494  &0.498  & 0.494\\
 &   & 0.20   & 0.193  & 0.192  &0.203  &0.196  &0.191 \\
  &   &0.05   & 0.052  &0.049   & 0.049 &0.052  & 0.049\\
~ & ~ & ~ & ~ & ~ & ~ & ~&~\\
&Equation (\ref{option two})& 0.95 &0.957 &0.974 &0.992 &0.975 &0.960\\
  &   & 0.80  &0.848 &0.865 &0.923 &0.868 &0.820\\
 &   &  0.50  &0.547 &0.573 &0.643 &0.561 &0.515 \\
 &   & 0.20   &0.224 &0.236 &0.274 &0.234 &0.204\\
  &   &0.05   &0.057 &0.059 &0.069 &0.056 &0.053\\
~ & ~ & ~ & ~ & ~ & ~ & ~&~\\
&$F$ pivot & 0.95 &0.901 &0.931 &0.940 &0.951 &0.947\\
  &   & 0.80  &0.750   &0.774   & 0.793 &0.794  &0.798 \\
 &   &  0.50  & 0.465  & 0.485  &0.492  &0.499  &0.499 \\
 &   & 0.20   & 0.189  &0.191   &0.203  & 0.199 &0.200 \\
  &   &0.05   &0.046   & 0.048  &0.049  &0.050  &0.047 \\
~ & ~ & ~ & ~ & ~ & ~ & ~&~\\
& Plug-in & 0.95 &0.886  &0.380 &0.729 &0.947 &0.947\\
  &   & 0.80  &0.728   & 0.256  & 0.538 &0.785  & 0.797\\
 &   &  0.50  & 0.446  &0.136   & 0.299 &0.491  &0.499 \\
 &   & 0.20   & 0.182  & 0.051  & 0.118 &0.195  &0.200 \\
  &   &0.05   &0.044   &0.011   &0.028  & 0.050 &0.047 \\
~ & ~ & ~ & ~ & ~ & ~ & ~&~\\
& Equation (\ref{option three}) & 0.95 &0.878  &0.944 &0.949 &0.946 &0.940\\
&$100p\%\equiv50\%$ & 0.80 &0.681  &0.787 &0.798 &0.788 &0.777\\
 &   &  0.50  &0.403  &0.481 &0.500 &0.497 &0.498\\
 &   & 0.20   &0.190  &0.185 &0.192 &0.200 &0.246\\
  &   &0.05  & 0.113 &0.040 &0.047 &0.068 &0.153\\
& & & & &\\
& Equation (\ref{option four}) & 0.95 &0.934  &0.939 & 0.945&0.948 &0.929\\
&$100p\%\equiv50\%$ & 0.80 & 0.791&0.789 &0.797 &0.804 &0.699\\
 &   &  0.50  &0.504  &0.495   &0.490  &0.495  &0.366 \\
 &   & 0.20   & 0.240  & 0.185  &0.201  &0.200  &0.151 \\
  &   &0.05   &  0.150 &0.043   &0.050  & 0.073 &0.097 \\
& & & & &\\
&Equation (\ref{option five})& 0.95 &0.953 &0.947 &0.955 &0.974 &0.984\\
 & $100p\%\equiv50\%$   & 0.80  &0.808 &0.796 &0.811 &0.851 &0.890\\
 &   &  0.50  &0.497 &0.491 &0.505 &0.547 &0.605\\
 &   & 0.20   &0.220 &0.192 &0.201 &0.217 &0.290\\
  &   &0.05   &0.120  &0.045 &0.047 &0.080 &0.162\\
~ & ~ & ~ & ~ & ~ & ~ & ~&~\\
\cline{1-1} \cline{2-2} \cline{3-3} \cline{4-4} \cline{5-5} \cline{6-6} \cline{7-7} \cline{8-8}
\end{tabular}
\end{center}
\caption{Estimated coverage probabilities under a gamma data process for study-level interarrival times over 10,000 Monte Carlo runs. The true coverage probability is considered maintained if the observed rate is within $\pm3$ standard errors, e.g. 0.943 to 0.957 for a true coverage probabiliy of 0.95.}
\label{table_sim_gamma_again}
\end{table}

\begin{table}[H]
\begin{center} 
\begin{tabular}{l l l r r r r r }
Data Generative  & Interval & Nom & n=10 & n=20 & n=100 & n=290 & n=299 \\
Process & & Cov& N=11& N=300 &N=300 & N=300 & N=300\\
& & Prob& & & & \\
\cline{1-1} \cline{2-2} \cline{3-3} \cline{4-4} \cline{5-5} \cline{6-6} \cline{7-7} \cline{8-8}
~ & ~ & ~ & ~ & ~ & ~ & ~&~\\
$Y_i\sim$ Gamma$\big(0.7,\frac{1.5}{0.7}\big)$ & Equation (\ref{option one}) & 0.95 & 0.857 &0.947 &0.947 &0.937 &0.835\\
  &   & 0.80  &0.704   & 0.787  & 0.795 &0.786  &0.702 \\
 &   &  0.50  &0.436   & 0.493  &0.491  &0.495  & 0.442\\
 &   & 0.20   &0.169   &0.191   &0.198  &0.193  &0.177 \\
  &   &0.05   &0.045   &0.051   &0.052  &0.051  & 0.045\\
~ & ~ & ~ & ~ & ~ & ~ & ~&~\\
&Equation (\ref{option two})& 0.95 &0.955 &0.978 &0.993 &0.975 &0.962\\
  &   & 0.80  &0.842 &0.866 &0.918 &0.861 &0.817\\
 &   &  0.50  &0.543 &0.577 &0.648 &0.564 &0.528\\
 &   & 0.20   &0.217 &0.232 &0.273 &0.229 &0.211\\
  &   &0.05   &0.058 &0.061 &0.071 &0.056 &0.052\\
~ & ~ & ~ & ~ & ~ & ~ & ~&~\\
&$F$ pivot & 0.95 &0.908 &0.938 &0.945 &0.951 &0.949\\
  &   & 0.80  &0.753   &0.774   &0.791  & 0.796 &0.793 \\
 &   &  0.50  & 0.462  & 0.490  &0.490  &0.497  &0.510 \\
 &   & 0.20   & 0.179  &0.195   &0.199  &0.199  &0.203 \\
  &   &0.05   &0.048   &0.049   & 0.051 &0.049  & 0.050\\
~ & ~ & ~ & ~ & ~ & ~ & ~&~\\
& Plug-in & 0.95 &0.892  &0.372 &0.733 &0.947 &0.949\\
  &   & 0.80  &0.724   & 0.251  &0.541  &0.789  &0.792 \\
 &   &  0.50  & 0.443  & 0.130  &0.297  &0.490  &0.509 \\
 &   & 0.20   & 0.172  &0.051   &0.119  &0.196  & 0.203\\
  &   &0.05   &0.045   &0.014   &0.029  &0.047  &0.050 \\
~ & ~ & ~ & ~ & ~ & ~ & ~&~\\
& Equation (\ref{option three}) & 0.95 &0.862  &0.942 &0.947 &0.944 &0.939\\
&$100p\%\equiv50\%$ & 0.80 & 0.672 &0.789 &0.804 &0.802 &0.775\\
 &   &  0.50  &0.397  &0.494 &0.496 &0.490 &0.483\\
 &   & 0.20   &0.193  &0.188 &0.196 &0.200 &0.252\\
  &   &0.05  &0.115  &0.047 &0.044 &0.070 &0.157\\
& & & & &\\
& Equation (\ref{option four}) & 0.95 &0.935  & 0.911&0.938 &0.960 &1.000\\
&$100p\%\equiv50\%$ & 0.80 &0.871 &0.781 &0.799 &0.832 &0.995\\
 &   &  0.50  & 0.716  & 0.497  &0.493  & 0.519 & 0.912\\
 &   & 0.20   &0.470   & 0.185  &0.200  & 0.205 &0.632 \\
  &   &0.05   & 0.332  & 0.045  &0.050  &0.076  & 0.446\\
& & & & &\\
&Equation (\ref{option five})& 0.95 &0.914 &0.948 &0.954 &0.970 &0.959\\
 & $100p\%\equiv50\%$   & 0.80  &0.749 &0.789 &0.808 &0.839 &0.834\\
 &   &  0.50  &0.449 &0.491 &0.505 &0.539 &0.548\\
 &   & 0.20   &0.202 &0.184 &0.203 &0.216 &0.268\\
  &   &0.05   &0.122 &0.043 &0.052 &0.075 &0.172\\
~ & ~ & ~ & ~ & ~ & ~ & ~&~\\
\cline{1-1} \cline{2-2} \cline{3-3} \cline{4-4} \cline{5-5} \cline{6-6} \cline{7-7} \cline{8-8}
\end{tabular}
\end{center}
\caption{Estimated coverage probabilities under a gamma data process for study-level interarrival times over 10,000 Monte Carlo runs. The true coverage probability is considered maintained if the observed rate is within $\pm3$ standard errors, e.g. 0.943 to 0.957 for a true coverage probabiliy of 0.95.}
\label{table_sim_gamma}
\end{table}

\section{Additional Figures}\label{additional figures}
\begin{figure}[H]
\centering
\includegraphics[trim={1.9cm 18cm 0 0}, clip, height = 2.2in]{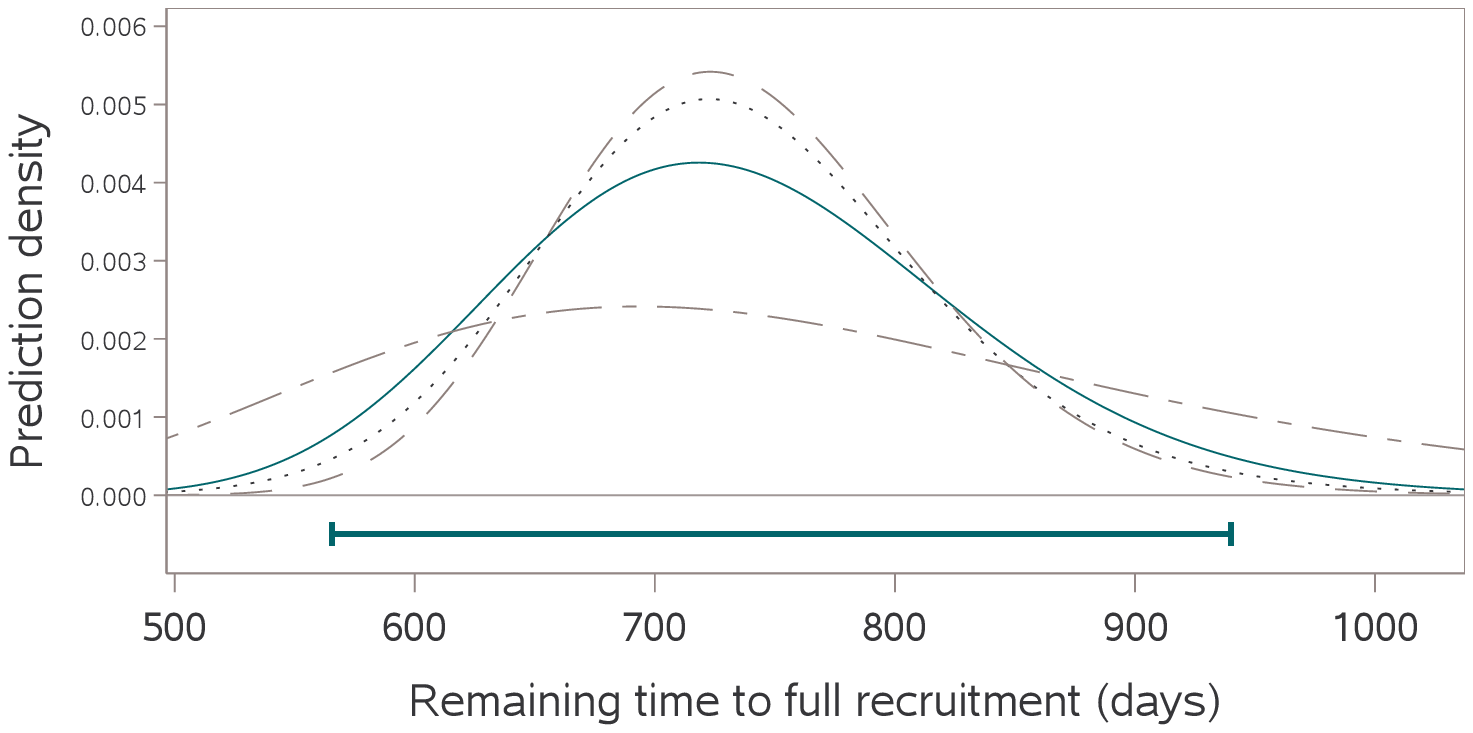}
\caption{\small{Solid density depicts prediction intervals of all levels for $\sum_{i=n+1}^{N} Y_i$, the remaining time to full recruitment using Equation (\ref{option two}). Error bars indicate the 95\% prediction limits. Dotted density shows prediction intervals corresponding to Equation (\ref{option one}). Dashed density shows prediction intervals corresponding to the pivotal quantity $\sum_{i=n+1}^{N} Y_i/(N-n)\bar{Y}_n\sim F\big(2(N-n)\hat{k},2n\hat{k}\big)$ treating $\hat{k}$ as known. Long-short density shows prediction intervals using the $F$ pivotal quantity while holding $k=1$.}} 
\label{prediction densities}
\end{figure}

\begin{figure}[H]
\centering
\includegraphics[trim={0.9cm 18cm 0 0}, clip, height = 2.2in]{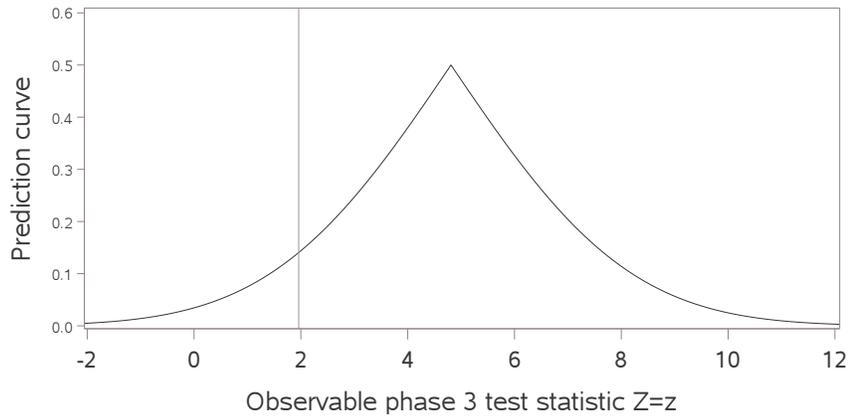}
\caption{\small{Solid curve shows p-values and prediction intervals corresponding to Equation (\ref{option one}), $[(\text{log}\{\hat{\rho}_n\}-0)/\hat{\text{se}}_n\sqrt{n/m} - z]/\sqrt{m/n+1}\overset{asymp}{\sim}T_{n-1}$.  The peak of the curve corresponds to the estimated phase 3 test statistic based on $n=100$ in phase 2.  The reference line identifies the value of the observed phase 3 test statistic needed to achieve statistical significance (1.96), and the height of the prediction confidence curve displays the p-value testing the hypothesis that the phase 3 result will be statistically significant at the one-sided 0.025 level.  P-value$=0.14$ testing $H_0$: $z\equiv[\text{log}\{\hat{\rho}_m\}-0]/\hat{\text{se}}_m \le 1.96$.}}
\label{prediction or z}
\end{figure}

\section{SAS Code}\label{sas code}
\footnotesize

\begin{lstlisting}

options nodate nonumber;
%let n=20; *Current enrollment;
%let total_enrollment=300; *Total planned enrollment;
%let additional_enrollment=%sysevalf(&total_enrollment.-&n.);

%let coverage=0.95;
%let alpha=%sysevalf(1-&coverage.);
%put &alpha.;



data gamma;
do i=1 to &total_enrollment.;
y=rand('gamma',4,2.5/4);
output;
end;
run;


options nodate nonumber;
ods graphics / border=no height=3in width=6.0in;
proc sgplot data=gamma;
where i le &n.;
histogram y;
xaxis label="Recruitment interarrival times (days)" values=(0 to 7 by 1)
 labelattrs=(size=14) valueattrs=(size=12);
yaxis labelattrs=(size=14) valueattrs=(size=12) offsetmax=0.1;
run;
  

  

proc genmod data=gamma;
where i le &n.;
model y = / dist=gamma link=identity lrci alpha=&alpha. covb;
estimate 'mu' int 1;
ods output parameterestimates=parms covb=covb;
run;

proc genmod data=gamma;
where i le &n.;
class i;
model y = / dist=gamma link=log alpha=&alpha.;
repeated subject=i / type=ind;
ods output GEEEmpPEst=parms_wald_log;
run;

proc sql noprint;

select estimate into: mu_hat from parms where parameter='Intercept';
select lowerLRCL into: mu_hat_lower from parms where parameter='Intercept';
select upperLRCL into: mu_hat_upper from parms where parameter='Intercept';
select stderr into: mu_se from parms where parameter='Intercept';
select estimate into: k_hat from parms where parameter='Scale';
select lowerLRCL into: k_hat_lower from parms where parameter='Scale';
select upperLRCL into: k_hat_upper from parms where parameter='Scale';
select stderr into:k_se from parms where parameter='Scale';
select scale into: covb from covb where rowname='Prm1';

select estimate into: log_est from parms_wald_log ;
select stderr into: se_log from parms_wald_log ;
select lowercl into: lower_log from parms_wald_log ;
select uppercl into: upper_log from parms_wald_log;
quit;
%put &mu_hat. &k_hat.;
%put &mu_hat_lower. &mu_hat_upper. &k_hat_lower. &k_hat_upper.;






data cd;
do mu=0 to 10 by 0.005;


mu_hat=&mu_hat.;

*Confidence distributions;

mu_se=&mu_se.; *model-based se;
H_wald=1-cdf('normal',(mu_hat-mu)/mu_se,0,1);
dH_wald_dmu=(H_wald-lag(H_wald))/(mu-lag(mu));

se_log=&se_log.; *sandwich se on log scale;
H_log_wald=1-cdf('normal',(log(mu_hat)-log(mu))/se_log,0,1);
dH_log_wald_dmu=(H_log_wald-lag(H_log_wald))/(mu-lag(mu));

nmu=&additional_enrollment.*mu;
dH_log_wald_dnmu=(H_log_wald-lag(H_log_wald))/(nmu-lag(nmu));


*Posterior (k known);
k_hat=&k_hat.;
n=&n.;
posterior=pdf('gamma',mu,(k_hat)*n,mu_hat/((k_hat)*n));




output;
end;
run;



data cd;
set cd;
lower_mu=exp(&lower_log.);
upper_mu=exp(&upper_log.);

additional_enrollment=&additional_enrollment.;

lower_nmu=&additional_enrollment.*lower_mu;
upper_nmu=&additional_enrollment.*upper_mu;

nmu_hat=&additional_enrollment.*mu_hat;
y_scatter=0;
run;




ods graphics / border=no height=3in width=6.0in;
ods escapechar="^";
proc sgplot data=cd noautolegend;
where 0.001 le H_log_wald le 0.999;
series x=mu y=dH_log_wald_dmu / lineattrs=(color="cx445694");
scatter x=mu_hat y=y_scatter / xerrorlower=lower_mu xerrorupper=upper_mu 
	markerattrs=(size=1) errorbarattrs=(color="cx445694" thickness=2) errorcapscale=0.5;

yaxis label="Confidence Density" labelattrs=(size=14) valueattrs=(size=9);
xaxis label="^{unicode mu}" labelattrs=(size=14) valueattrs=(size=9);
run;


options nodate nonumber;
ods graphics / border=no;
ods escapechar="^";
proc sgplot data=cd noautolegend;
where 0.001 le H_log_wald le 0.999;
series x=nmu y=dH_log_wald_dnmu / lineattrs=(color="cx445694");

scatter x=nmu_hat y=y_scatter / xerrorlower=lower_nmu xerrorupper=upper_nmu 
	markerattrs=(size=1) errorbarattrs=(color="cx445694" thickness=2) errorcapscale=0.5;

yaxis label="Confidence Density" labelattrs=(size=14) valueattrs=(size=9);
xaxis label="Mean time to full recruitment (days)" labelattrs=(size=14) 
valueattrs=(size=12) offsetmax=0.07;*label="(N-n) ^{unicode mu}";
run;
  

data sum_y;
do sum_y=0 to 3000 by 0.5;
pdf_lower=pdf('gamma',sum_y,&additional_enrollment.*&k_hat.,exp(&lower_log.)/&k_hat.);
pdf_upper=pdf('gamma',sum_y,&additional_enrollment.*&k_hat.,exp(&upper_log.)/&k_hat.);
cdf_lower=cdf('gamma',sum_y,&additional_enrollment.*&k_hat.,exp(&lower_log.)/&k_hat.);
cdf_upper=cdf('gamma',sum_y,&additional_enrollment.*&k_hat.,exp(&upper_log.)/&k_hat.);



pred_lower=quantile('gamma',0.025,&additional_enrollment.*&k_hat.,exp(&lower_log.)/&k_hat.);
pred_upper=quantile('gamma',0.975,&additional_enrollment.*&k_hat.,exp(&upper_log.)/&k_hat.);
call symput('pred_lower',pred_lower);
call symput('pred_upper',pred_upper);

additional_enrollment=&additional_enrollment.;
mu_hat=&mu_hat.;
mu_lower=exp(&lower_log.);
mu_upper=exp(&upper_log.);
nmu_hat=&additional_enrollment.*&mu_hat.;
nmu_hat_lower=&additional_enrollment.*exp(&lower_log.);
nmu_hat_upper=&additional_enrollment.*exp(&upper_log.);


y_scatter=0.0006;
y_scatter2=-0.00035;



output;
end;
run;

data sum_y2;
set sum_y;


if cdf_lower gt 0.99999 then pdf_lower=.; if cdf_lower2 gt 0.99999 then pdf_lower2=.;
if cdf_upper lt 0.00001 then pdf_upper=.; if cdf_upper2 lt 0.00001 then pdf_upper2=.;

if cdf_lower lt 0.00001 then pdf_lower=.; if cdf_lower2 lt 0.00001 then pdf_lower2=.;
if cdf_upper gt 0.99999 then pdf_upper=.; if cdf_upper2 gt 0.99999 then pdf_lower2=.;

if 0.49 gt cdf_lower gt 0.51  then nmu_hat=.;

run;


*pred;
options nodate nonumber;
proc sgplot data=sum_y2 noautolegend;
where cdf_lower gt 0.00001 and cdf_upper lt 0.99999;
series x=sum_y y=pdf_lower / lineattrs=(color="cxA23A2E");
series x=sum_y y=pdf_upper / lineattrs=(color="cxA23A2E");


refline nmu_hat_lower / axis=x;
refline nmu_hat_upper / axis=x;
scatter x=nmu_hat y=y_scatter / xerrorlower=pred_lower xerrorupper=pred_upper
	markerattrs=(size=1) errorbarattrs=(color="cx01665E" thickness=2) errorcapscale=0.5;
scatter x=nmu_hat y=y_scatter2 / xerrorlower=nmu_hat_lower xerrorupper=nmu_hat_upper
	markerattrs=(size=1) errorbarattrs=(color="cx445694" thickness=2) errorcapscale=0.5 ;

xaxis  label="Remaining time to full recruitment (days)"   
labelattrs=(size=14) valueattrs=(size=12); 
yaxis label="Sampling Density" labelattrs=(size=14) valueattrs=(size=9);
run;
  




data prediction;
set cd;

pred_quantile=quantile('gamma',H_log_wald,&additional_enrollment.*&k_hat.,mu/&k_hat.);
pred_pdf=(H_log_wald-lag(H_log_wald))/(pred_quantile-lag(pred_quantile));
pred_cdf=H_log_wald;
pred_curve=pred_cdf*(pred_quantile<&additional_enrollment.*exp(&log_est.)) 
	+ (1-pred_cdf)*(pred_quantile>&additional_enrollment.*exp(&log_est.));



pred_t_cdf=1-cdf('t',(&log_est.-log(pred_quantile/(&total_enrollment.-&n.)))/
	(sqrt(&n.)*&se_log.*sqrt(1/&n. + 1/(&total_enrollment.-&n.))),&n.-1);
pred_t_pdf=(pred_t_cdf-lag(pred_t_cdf))/(pred_quantile-lag(pred_quantile));
pred_t_curve=pred_t_cdf*(pred_quantile<&additional_enrollment.*exp(&log_est.)) 
	+ (1-pred_t_cdf)*(pred_quantile>&additional_enrollment.*exp(&log_est.));


pred_F_cdf=cdf('F',pred_quantile/(&additional_enrollment.*&mu_hat.),2*(&additional_enrollment.)
	*&k_hat.,2*&n.*&k_hat.);
pred_F_pdf=(pred_F_cdf-lag(pred_F_cdf))/(pred_quantile-lag(pred_quantile));
pred_F_curve=pred_F_cdf*(pred_quantile<&additional_enrollment.*exp(&log_est.)) 
	+ (1-pred_F_cdf)*(pred_quantile>&additional_enrollment.*exp(&log_est.));

pred_F1_cdf=cdf('F',pred_quantile/(&additional_enrollment.*&mu_hat.),2*(&additional_enrollment.)*1,
	2*&n.*1);
pred_F1_pdf=(pred_F1_cdf-lag(pred_F1_cdf))/(pred_quantile-lag(pred_quantile));
pred_F1_curve=pred_F1_cdf*(pred_quantile<&additional_enrollment.*exp(&log_est.)) 
	+ (1-pred_F1_cdf)*(pred_quantile>&additional_enrollment.*exp(&log_est.));
if pred_quantile < &additional_enrollment.*exp(&log_est.) then pred_F1_curve_lower=pred_F1_cdf;
else if pred_quantile ge &additional_enrollment.*exp(&log_est.) then pred_F1_curve_lower=.;
if pred_quantile > &additional_enrollment.*exp(&log_est.) then pred_F1_curve_upper=1-pred_F1_cdf;
else if pred_quantile le &additional_enrollment.*exp(&log_est.) then pred_F1_curve_upper=.;



pred_f1_lower=&additional_enrollment.*&mu_hat.
	*quantile('f',&alpha./2,2*&additional_enrollment.,2*&n.);
pred_f1_upper=&additional_enrollment.*&mu_hat.
	*quantile('f',1-&alpha./2,2*&additional_enrollment.,2*&n.);





pred_lower=&pred_lower.;
pred_upper=&pred_upper.;
nmu_hat=&additional_enrollment.*&mu_hat.;

y_scatter=0.000025;


run;


proc sgplot data=prediction noautolegend;
where 0.001 le pred_cdf le 0.999;
series x=pred_quantile y=pred_t_curve / lineattrs=(color=black pattern=dot);;
series x=pred_quantile y=pred_F_curve / lineattrs=(color=grey pattern=dash) name="fpivot"
	 legendlabel="F-pivot";
series x=pred_quantile y=pred_F1_curve_lower / lineattrs=(color=grey pattern=8);
series x=pred_quantile y=pred_F1_curve_upper / lineattrs=(color=grey pattern=8);
series x=pred_quantile y=pred_curve / lineattrs=(color="cx01665E")  name="eq1" 
	legendlabel="Equation (2)";
yaxis max=0.6;
scatter y=y_scatter x=nmu_hat / xerrorlower=pred_lower xerrorupper=pred_upper
	markerattrs=(size=1) errorbarattrs=(color="cx01665E" thickness=2) errorcapscale=0.5;
xaxis label="Remaining time to full recruitment (days)"  
labelattrs=(size=14) valueattrs=(size=12);
yaxis label="Prediction Curve" labelattrs=(size=14) valueattrs=(size=9);
run;


options nodate nonumber;
proc sgplot data=prediction noautolegend;
where 0.001 le pred_cdf le 0.999;
series x=pred_quantile y=pred_pdf / lineattrs=(color="cx01665E")  name="eq1" 
	legendlabel="Equation (2)";
series x=pred_quantile y=pred_t_pdf / lineattrs=(color=black pattern=dot);
series x=pred_quantile y=pred_F_pdf / lineattrs=(color=grey pattern=dash) name="fpivot" 
	legendlabel="F-pivot";
series x=pred_quantile y=pred_F1_pdf / lineattrs=(color=grey pattern=8);
scatter y=y_scatter x=nmu_hat / xerrorlower=pred_lower xerrorupper=pred_upper
	markerattrs=(size=1) errorbarattrs=(color="cx01665E" thickness=2) errorcapscale=0.5;
xaxis label="Remaining time to full recruitment (days)"  
labelattrs=(size=14) valueattrs=(size=12); 
yaxis label="Prediction Density" labelattrs=(size=14) valueattrs=(size=9);
run;
  
  
  
*Tolerance;

data tolerance;
do quantile=100 to 1500 by 1;

cdf=cdf('gamma',quantile,&total_enrollment.*&k_hat.,&mu_hat./&k_hat.);
pdf=pdf('gamma',quantile,&total_enrollment.*&k_hat.,&mu_hat./&k_hat.);


tol_quantile_hat_05=quantile('gamma',0.05,&total_enrollment.*&k_hat.,&mu_hat./&k_hat.);
tol_quantile_hat_95=quantile('gamma',0.95,&total_enrollment.*&k_hat.,&mu_hat./&k_hat.);

deriv_mu_05=( quantile('gamma',0.05,&total_enrollment.*&k_hat.,&mu_hat./&k_hat.)
	-quantile('gamma',0.05,&total_enrollment.*&k_hat.,(&mu_hat.-0.001)/&k_hat.) )/0.001;
deriv_k_05=( quantile('gamma',0.05,&total_enrollment.*&k_hat.,&mu_hat./&k_hat.)
 -quantile('gamma',0.05,&total_enrollment.*(&k_hat.-0.001),(&mu_hat.)/(&k_hat.-0.001) ) )/0.001;
se_05=sqrt( (deriv_mu_05*&mu_se.)**2 + (deriv_k_05*&k_se.)**2 + 2*deriv_mu_05*deriv_k_05*&covb. );

deriv_mu_95=( quantile('gamma',0.95,&total_enrollment.*&k_hat.,&mu_hat./&k_hat.)
	-quantile('gamma',0.95,&total_enrollment.*&k_hat.,(&mu_hat.-0.001)/&k_hat.) )/0.001;
deriv_k_95=( quantile('gamma',0.95,&total_enrollment.*&k_hat.,&mu_hat./&k_hat.)
 -quantile('gamma',0.95,&total_enrollment.*(&k_hat.-0.001),(&mu_hat.)/(&k_hat.-0.001) ) )/0.001;
se_95=sqrt( (deriv_mu_95*&mu_se.)**2 + (deriv_k_95*&k_se.)**2 + 2*deriv_mu_95*deriv_k_95*&covb. );

H=1-cdf('normal',(log(tol_quantile_hat_05)-log(quantile))/(se_05/tol_quantile_hat_05),0,1 );
H_minus=cdf('normal',(log(tol_quantile_hat_95)-log(quantile))/(se_95/tol_quantile_hat_95),0,1 );

if quantile lt tol_quantile_hat_05 then C_lower=H; else C_lower=.;
if quantile gt tol_quantile_hat_95 then C_upper=H_minus; else C_upper=.;

output;
end;
run;

proc sql noprint;
select max(quantile)
into: lower_limit
from tolerance
where . lt C_lower le 0.025;
select max(quantile)
into: lower_point
from tolerance
where . lt C_lower;

select min(quantile)
into: upper_limit
from tolerance
where . lt C_upper le 0.025;
select min(quantile)
into: upper_point
from tolerance
where . lt C_upper;
quit;

%put &lower_limit. &upper_limit. &lower_point. &upper_point.;

data tolerance;
set tolerance;
lower_limit=&lower_limit.;
upper_limit=&upper_limit.;
 y_scatter=-0.00025;
lower_point=&lower_point.;
upper_point=&upper_point.;
x_scatter=750;
if pdf lt 0.00005 then pdf=.;
run;





options nodate nonumber;
proc sgplot data=tolerance noautolegend;
where (~(H lt 0.01) and ~(H_minus lt 0.01)) or pdf gt .;
series x=quantile y=C_lower / y2axis;
series x=quantile y=C_upper / y2axis;

refline lower_point  / axis=x;
refline upper_point / axis=x;
series x=quantile y=pdf / lineattrs=(color=cxA23A2E);
xaxis label="Total time to full recruitment (days)"  
	valueattrs=(size=14) labelattrs=(size=14);
yaxis label="Estimated Density" valueattrs=(size=9) labelattrs=(size=14);
y2axis max=1 label="Tolerance Curve" valueattrs=(size=9) labelattrs=(size=14);

scatter x=x_scatter y=y_scatter / markerattrs=(size=0.01) 
xerrorlower=lower_limit xerrorupper=upper_limit 
errorbarattrs=(color=cx445694 thickness=2) errorcapscale=0.5;
run;





*Poisson or negative binomial modeling;

data gamma2;
set gamma;
event=1;
run;


data gamma2;
set gamma2;
by event;
retain keep;
if first.event then keep=y;
else if not first.event then do;
keep=y+keep;
end;
if keep le 7 then week=1;
if 7 lt keep le 14 then week=2;
if 14 lt keep le 21 then week=3;
if 21 lt keep le 28 then week=4;
if 28 lt keep le 35 then week=5;
if 35 lt keep le 42 then week=6;
if 42 lt keep le 49 then week=7;
if 49 lt keep le 56 then week=8;
run;

proc means data=gamma2 sum noprint;
where i le &n.;
class week;
var event;
output out=counts (where=(week ne .)) sum(event)=events;
run;

ods graphics / border=no height=3in width=6.0in;
options nodate nonumber;

proc sgplot data=counts;
histogram events ;
xaxis label="Number of subjects recruited per week" min=0 labelattrs=(size=14) 
valueattrs=(size=12) values=(0 to 6 by 1);
yaxis label="Percent" offsetmax=0.05 labelattrs=(size=14) valueattrs=(size=12);			
run;

  

proc means data=counts mean var;
var events;
run;




data gamma;
set gamma;
event=1;
offset=y/(7);*7;
*offset=y;
log_offset=log(offset);
rate=event/offset;
run;



proc genmod data=gamma ;
where i le &n.;
model event = / dist=poisson link=log offset=log_offset  scale=deviance;
*model event = / dist=negbin link=log offset=log_offset noscale scale=0;
estimate 'lambda' int 1 / exp;
ods output parameterestimates=count_parms;
run;




data count_parms;
set count_parms;
if parameter='Intercept' then do;


estimate=exp(estimate);
lowerwaldcl=exp(lowerwaldcl);
upperwaldcl=exp(upperwaldcl);
end;
run;



proc sql noprint;
select sum(events) into: sum_events from counts;
select sum(y) into: offset from gamma where i le &n.;
select estimate into: lambda_hat from count_parms where parameter='Intercept';
select lowerwaldCL into: lambda_hat_lower_wald from count_parms where parameter='Intercept';
select upperwaldCL into: lambda_hat_upper_wald from count_parms where parameter='Intercept';
select stderr into: stderr from count_parms where parameter='Intercept';

select estimate into: phi_hat from count_parms where parameter='Scale';
quit;
%put  &lambda_hat. &phi_hat. &offset.;
%put &lambda_hat_lower_wald. &lambda_hat_upper_wald. ;





data cd_lambda;
do lambda=0.01 to 6 by 0.002;


phi_hat=&phi_hat.;
lambda_hat=&lambda_hat.;

sum_events=&sum_events.;
offset=&offset./7;




*transformed Wald on linear predictor from Genmod with underdispersion;
lambda_hat_lower=&lambda_hat_lower_wald.;
lambda_hat_upper=&lambda_hat_upper_wald.;
stderr=&stderr.;

z=(log(lambda_hat)-log(lambda))/stderr;

H=1-cdf('normal',z,0,1);

dH_dlambda=(H-lag(H))/(lambda-lag(lambda));







output;
end;
run;



data cd_lambda;
set cd_lambda;


w104lambda=104.29*lambda;
dH_dw104lambda=(H-lag(H))/(w104lambda-lag(w104lambda));
lambda_hat_104=104.29*lambda_hat;
lambda_hat_lower_104=lambda_hat_lower*104.29;
lambda_hat_upper_104=lambda_hat_upper*104.29;
y_scatter=0;
run;


options nodate nonumber;
ods graphics / border=no;
ods escapechar="^";
proc sgplot data=cd_lambda noautolegend;
where 0.001 le H le 0.999;
series x=lambda y=dH_dlambda;
yaxis offsetmax=0.05 label="Confidence Density" labelattrs=(size=14) valueattrs=(size=9);
scatter x=lambda_hat y=y_scatter / xerrorlower=lambda_hat_lower xerrorupper=lambda_hat_upper
	markerattrs=(size=1) errorbarattrs=(color="cx445694" thickness=2) errorcapscale=0.5;
xaxis label="Mean number of subjects recruited per week" labelattrs=(size=14) 
valueattrs=(size=12)  ;
run;
  

options nodate nonumber;
ods graphics / border=no;
proc sgplot data=cd_lambda noautolegend;
where 0.001 le H le 0.999;
series x=w104lambda y=dH_dw104lambda;
yaxis offsetmax=0.05 label="Confidence Density" labelattrs=(size=14) valueattrs=(size=9);
scatter x=lambda_hat_104 y=y_scatter / xerrorlower=lambda_hat_lower_104 
	xerrorupper=lambda_hat_upper_104 markerattrs=(size=1) 
	errorbarattrs=(color="cx445694" thickness=2) errorcapscale=0.5;
xaxis label="Mean number of subjects recruited per 730 days" labelattrs=(size=14) 
valueattrs=(size=12) ;
run;
  



*---------;
data gamma;
set gamma;
event=1;
offset=y/(7*104.29);
*offset=y;
log_offset=log(offset);
rate=event/offset;
run;



proc genmod data=gamma ;
where i le &n.;
model event = / dist=poisson link=log offset=log_offset  scale=deviance;
*model event = / dist=negbin link=log offset=log_offset noscale scale=0;
estimate 'lambda' int 1 / exp;
ods output parameterestimates=count_parms;
run;

proc sql noprint;
select estimate into: log_est_count from count_parms where parameter='Intercept';
select stderr into: se_log_count from count_parms where parameter='Intercept';
quit;



*----------;


data prediction_counts;
set cd_lambda;

pred_quantile=quantile('poisson',H,104.29*lambda);
*Normal approximation with underdispersion;
pred_quantile=quantile('normal',H,104.29*lambda,sqrt(104.29*lambda*phi_hat)); 
*Gamma approximation with underdispersion;
pred_quantile=quantile('gamma',H,104.29*lambda/phi_hat,phi_hat); 
pred_pdf=(H-lag(H))/(pred_quantile-lag(pred_quantile));
pred_cdf=H;
pred_curve=pred_cdf*(pred_quantile<exp(&log_est_count.)) 
	+ (1-pred_cdf)*(pred_quantile>exp(&log_est_count.));



if 0.0245 le H le 0.0255 then pred_lower=pred_quantile;
if 0.9745 le H le 0.9755 then pred_upper=pred_quantile;
lambda_hat104=lambda_hat*104.29;

y_scatter=0;

pred_cdf_t=1-cdf('normal',(&log_est_count.-log(pred_quantile))/
	sqrt(&se_log_count.**2 +  &se_log_count.**2) ,0,1);
pred_pdf_t=(pred_cdf_t-lag(pred_cdf_t))/(pred_quantile-lag(pred_quantile));
pred_curve_t=pred_cdf_t*(pred_quantile<exp(&log_est_count.)) 
	+ (1-pred_cdf_t)*(pred_quantile>exp(&log_est_count.));

pred_cdf_kris=1-cdf('normal',(104.29*lambda_hat*offset - offset*pred_quantile)/
	sqrt(phi_hat*(104.29*offset*(lambda_hat*offset + pred_quantile))),0,1);
pred_pdf_kris=(pred_cdf_kris-lag(pred_cdf_kris))/(pred_quantile-lag(pred_quantile));
pred_curve_kris=pred_cdf_kris*(pred_quantile<exp(&log_est_count.)) 
	+ (1-pred_cdf_kris)*(pred_quantile>exp(&log_est_count.));

run;


proc sql noprint;
select median(pred_quantile) into: pred_lower from prediction_counts 
where  0.0245 le H le 0.0255;

select median(pred_quantile) into: pred_upper from prediction_counts 
where  0.9745 le H le 0.9755;
quit;


data prediction_counts;
set prediction_counts;
pred_lower=&pred_lower.;
pred_upper=&pred_upper.;
run;



options nodate nonumber;
proc sgplot data=prediction_counts noautolegend;
where 0.001 le pred_cdf le 0.999;
series x=pred_quantile y=pred_curve / lineattrs=(color="cx01665E");
series x=pred_quantile y=pred_curve_kris / lineattrs=(pattern=dash color=grey);
series x=pred_quantile y=pred_curve_t / lineattrs=(pattern=dot color=black);
scatter y=y_scatter x=lambda_hat_104 / xerrorlower=pred_lower xerrorupper=pred_upper
	markerattrs=(size=1) errorbarattrs=(color="cx01665E" thickness=2) errorcapscale=0.5;
xaxis label="Number of additional subjects recruited in the next 730 days"  labelattrs=(size=14) 
valueattrs=(size=12) ;
yaxis label="Prediction Curve" max=0.5 labelattrs=(size=14) valueattrs=(size=9);
run;
  

options nodate nonumber;
proc sgplot data=prediction_counts noautolegend;
where 0.001 le pred_cdf le 0.999;
series x=pred_quantile y=pred_pdf / lineattrs=(color="cx01665E");
series x=pred_quantile y=pred_pdf_kris / lineattrs=(pattern=dash color=grey);
series x=pred_quantile y=pred_pdf_t / lineattrs=(pattern=dot color=black);
scatter y=y_scatter x=lambda_hat_104 / xerrorlower=pred_lower xerrorupper=pred_upper
	markerattrs=(size=1) errorbarattrs=(color="cx01665E" thickness=2) errorcapscale=0.5;
xaxis label="Number of additional subjects recruited in the next 730 days"  labelattrs=(size=14) 
valueattrs=(size=12) ;
yaxis label="Prediction Density" offsetmax=0.05 labelattrs=(size=14) valueattrs=(size=9);
run;




/* Clinical Trial Success Example*/

options nodate nonumber;

data bin;
do i=1 to 100;
	if i le 60 then do;
		trt=1;
		y=rand('bernoulli',0.25); 
	end;
	else do;
		trt=0;
		y=rand('bernoulli',0.1);
	end;
output;
end;
run;


proc means data=bin mean;
class trt;
var y;
run;

proc genmod data=bin descending;
class trt (ref='0');
model y = trt/ dist=bin link=logit;
estimate 'trt=0' int 1 trt 0 1 ;
estimate 'trt=1' int 1 trt 1 0 ;
estimate 'OR' trt 1 -1 / exp;
ods output parameterestimates=parms (where=(level1='1'));
run;

data parms;
set parms;
call symput('estimate',estimate);
call symput('stderr',stderr);
run;





data phase3;
do or=0 to 100 by 0.05;
log_or2_hat=&estimate.;
se_log_or3_hat=sqrt(100)*&stderr./sqrt(600);
log_or3_hat=1.96*se_log_or3_hat;

phase3_power=1-cdf('normal',(log_or3_hat-log(or))/se_log_or3_hat,0,1 );

call symput('ref',exp(log_or3_hat));

or3_hat=exp(log_or3_hat);

call symput('mde',or3_hat);

H=1-cdf('normal',(&estimate.-log(or))/&stderr. ,0,1);
dH_dor=(H-lag(H))/(or-lag(or));
C=H*(log(or)<&estimate.) + (1-H)*(log(or)>&estimate.);

weight=(or-lag(or))*dH_dor;

output;
end;
run;

ods graphics / border=no height=3in width=6.0in;
proc sgplot data=phase3 noautolegend;
where H le 0.99;
series x=or y=C / y2axis name="cc" legendlabel="Confidence Curve";
series x=or y=phase3_power / lineattrs=(thickness=2) name="phase3_power" 
	legendlabel="Phase 3 Power";
y2axis max=0.6 label="Confidence Curve" labelattrs=(size=14) valueattrs=(size=9);
yaxis label="Phase 3 Power" labelattrs=(size=14) valueattrs=(size=9);
xaxis label="True Population-Level Odds Ratio" labelattrs=(size=14) valueattrs=(size=12);
keylegend "phase3_power" "cc" / location=inside;
run;



proc sgplot data=phase3 noautolegend;
where H le 0.99;
series x=or y=dH_dor / y2axis name="cd" legendlabel="Confidence Density";
series x=or y=phase3_power / lineattrs=(thickness=2) name="phase3_power" 
	legendlabel="Phase 3 Power";
y2axis offsetmax=0.2 label="Confidence Density" labelattrs=(size=14) valueattrs=(size=9);
yaxis label="Phase 3 Power" labelattrs=(size=14) valueattrs=(size=9);
xaxis label="True Population-Level Odds Ratio" labelattrs=(size=14) valueattrs=(size=12);
keylegend "phase3_power" "cd" / location=inside;
run;


data parms2;
set parms;
do or3_hat=0 to 100 by 0.05;

H=1-cdf('normal',(estimate-log(or3_hat))/(sqrt(100)*stderr*sqrt(1/100 + 1/600)),0,1  );
dH_dor3_hat=(H-lag(H))/(or3_hat-lag(or3_hat));
C=H*(or3_hat<exp(estimate)) + (1-H)*(or3_hat>exp(estimate));


output;
end;
run;



options nodate nonumber;
proc sgplot data=parms2;
where H le 0.99;
series x=or3_hat y=C / lineattrs=(pattern=solid color=black thickness=1);
refline &ref./ axis=x;
xaxis label="Observed phase 3 odds ratio" labelattrs=(size=14) valueattrs=(size=12);
yaxis label="Prediction Curve" max=0.6 labelattrs=(size=14) valueattrs=(size=9);
run;


proc sql;
title 'Phase 3 MDE';
select distinct or3_hat
from phase3;

title 'p-value testing H0: phase3 or <= mde';
select max(H)
from parms2
where or3_hat le &mde.;
quit;

title;
proc means data=phase3 sum;
var weight;
run;

title 'PoS';
proc means data=phase3 mean;
weight weight;
var phase3_power;
run;
title;



  



\end{lstlisting}

\end{document}